\documentclass[modern]{aastex61}

\received{October, 2018}
\revised{February, 2019}
\submitjournal{ApJ}

\shorttitle{A spectroscopic Analysis of the CKS}
\shortauthors{Martinez et al.}

\begin{document}

\title{A Spectroscopic Analysis of the California-$Kepler$ Survey Sample: I. Stellar Parameters, Planetary Radii and a Slope in the Radius Gap}

\correspondingauthor{Cintia F. Martinez}
\email{cmartinez@on.br}

\author[0000-0002-0786-7307]{Cintia F. Martinez}
\affiliation{Observat\'orio Nacional, Rua General Jos\'e Cristino, 77, 20921-400 S\~ao Crist\'ov\~ao, Rio de Janeiro, RJ, Brazil}

\author{Katia Cunha}
\affiliation{Observat\'orio Nacional, Rua General Jos\'e Cristino, 77, 20921-400 S\~ao Crist\'ov\~ao, Rio de Janeiro, RJ, Brazil}
\affiliation{Steward Observatory, University of Arizona, 933 North Cherry Ave., Tucson, AZ 85721, USA}

\author{Luan Ghezzi}
\affiliation{Observat\'orio Nacional, Rua General Jos\'e Cristino, 77, 20921-400 S\~ao Crist\'ov\~ao, Rio de Janeiro, RJ, Brazil}

\author{Verne V. Smith}
\affiliation{National Optical Astronomy Observatory, 950 North Cherry Avenue, Tucson, AZ 85719, USA}

\begin{abstract}

We present results from a quantitative spectroscopic analysis conducted on archival Keck/HIRES high-resolution spectra from the California-$Kepler$ Survey (CKS) sample of transiting planetary host stars identified from the $Kepler$ mission. The spectroscopic analysis was based on a carefully selected set of Fe I and Fe II lines, resulting in precise values for the stellar parameters of effective temperature (T$_{\rm eff}$) and surface gravity (log $g$).
Combining the stellar parameters with $Gaia$ DR2 parallaxes and precise distances, we derived both stellar and planetary radii for our sample, with a median internal uncertainty of 2.8$\%$ in the stellar radii and 3.7$\%$ in the planetary radii. 
An investigation into the distribution of planetary radii confirmed the bimodal nature of this distribution for the small radius planets found in previous studies, with peaks at: $\sim$1.47 $\pm$ 0.05 R$_{\oplus}$ and $\sim$2.72 $\pm$ 0.10 R$_{\oplus}$, with a gap at $\sim$ 1.9R$_{\oplus}$. 
Previous studies that modeled planetary formation that is dominated by photo-evaporation predicted this bimodal radii distribution and the presence of a radius gap, or photo-evaporation valley.  Our results are in overall agreement with these models.
The high internal precision achieved here in the derived planetary radii clearly reveal the presence of a slope in the photo-evaporation valley for the CKS sample, indicating that the position of the radius gap decreases with orbital period; this decrease was fit by a power law of the form R$_{pl}$ $\propto$ P$^{-0.11}$, which is consistent with photo-evaporation and Earth-like core composition models of planet formation.

\end{abstract}


\section{Introduction} \label{sec:intro}

One well-known axiom in exoplanetary studies connects the properties of exoplanets with the properties of their host stars through the expression that ``one can only know the planet to the level that the host star is known''.  
To know certain exoplanetary physical properties (such as radius, mass, and mean density) requires the knowledge of those same physical properties for the host star: for transiting exoplanets, it is possible to determine the planetary radius relative to the stellar 
radius (R$_{p}$/R$_{\star}$) from the analysis of the transit light curve, while the planetary mass depends on the host stellar mass 
(M$_{p}$ $\propto$ M$_{\star}^{2/3}$) and is derived from the radial velocity curve.  

Although several thousand exoplanet candidates have been discovered by the $Kepler$ mission \citep{borucki2010,koch2010,borucki2016}, the initial
stellar pa\-ra\-meters, as derived from the Kepler Input Catalog (KIC), were limited in accuracy, as the KIC provided 7-band photometry (g, r, i, and z, plus J, H, and K from 2MASS), along with a narrow filter centered on the Mg I b-lines used as a luminosity indicator \citep{borucki2016}. 
The deduced stellar radii were found to have a scatter of about 30-40$\%$ \citep{huber2014}, with the errors for late-type dwarfs being even larger \citep{dressing2013}. Errors of this size result in large uncertainties in exo\-pla\-netary properties and
could mask correlations and trends in exoplanet properties or the types of systems they inhabit.

Improvements in the derived stellar radii can be accomplished by conducting precise, quantitative spectroscopic analyses of the host stars, in particular, using high-quality, high-resolution spectra. In addition to high-resolution spectroscopic data, $Gaia$ 
\citep{gaia2018} has now provided precise parallaxes for
a large number of $Kepler$ exoplanet host-stars. The $Gaia$ parallaxes, combined with tightly-constrained stellar parameters derived from spectroscopy, result in stellar and exoplanetary radii with accuracies of 3-5$\%$ \citep{stassun2017}.

A recent example of how improved measurements for stellar radii can reveal new characteristics in exoplanet populations can be found from the California-$Kepler$ Survey \citep[CKS-][]{petigura2017}, where \cite{fulton2017} discovered a bimodal distribution for
small planet radii, with peaks at $\sim$1.3R$_{\oplus}$ and $\sim$2.4R$_{\oplus}$, and a gap in-between that points to a transition radius separating super-Earths from sub-Neptunes.
An earlier detection of this small-planet gap was prevented due to the
large uncertainties in the exoplanetary radii, although
a small-planet gap was originally predicted by several formation models 
\citep{owen2013,lopez2014,jin2014,chen2016,lopez2016}, which predicted
that gaseous planets may suffer photo-evaporation of their envelopes by
radiation coming from their host stars. The presence of the small-planet gap has now been confirmed by other studies \citep{vaneylen2018,fulton2018,berger2018}. 
In addi\-tion to photo-evaporation, \cite{ginzburg2016} and more recently, \cite{ginzburg2018} have shown that the small-planet gap can also be produced by a young, hot planetary core, whose energy can drive atmospheric mass loss, with the ability to retain an atmosphere depending on the mass of the planet.  Other processes, such as internal planetary outgassing \citep{dorn2018}, or large impacts on young planets \citep{inamdar2016} can both produce a planetary atmosphere or remove it.

In this study, a homogeneous spectroscopic analysis has been carried out in order to derive precise stellar parameters (effective temperatures, T$_{\rm eff}$ and surface gravity) for a sample of $Kepler$ hosts, using a homogeneous set of high-resolution optical Keck/HIRES spectra made public by the California-$Kepler$ Survey team \citep{petigura2017,johnson2017,fulton2017}. Stellar radii are derived using the $Gaia$ DR2 parallaxes \citep{bailer2018,lindegren2018}, with the improvements in the stellar radii given by the precise parallaxes and better distances from $Gaia$. Based on our analysis, we independently derive planetary radii and confirm the presence of a small-planet gap in the distribution of exoplanetary sizes.

\section{Observations} \label{sec:obser}

The high-resolution spectra analyzed in this study were obtained as part of the California-$Kepler$ Survey \citep[CKS-][]{petigura2017,johnson2017,fulton2017}, a large observational campaign targeting stars identified as $Kepler$ Objects of Interest (KOI's). 
The CKS campaign was conducted between 2012 - 2014 using the High Resolution Echelle Spectrometer \citep[HIRES-][]{vogt1994} at the Keck telescope.
All CKS spectra analyzed here were reduced by \cite{petigura2017} and are publicly available on the Keck Observatory Archive. The spectra were obtained from https://california-planet-search.github.io/cks-website/. (Sample HIRES spectra of the CKS are shown in Figure 2 from \citealt{petigura2017}).

From the full CKS sample of 1305 stars, we removed a small sample of 20 stars that had low quality spectra (all having S/N ratios lower than $\sim$30).
The remaining sample containing 1285 stars was analyzed spec\-tros\-co\-pi\-cally in this study; most of these stars have spectra with S/N ratios between 40-70, while some $\sim$ 8$\%$ of the spectra have excellent quality with S/N higher than 100.

\section{Analysis} \label{sec:method}

\subsection{Spectroscopic Stellar Parameters} \label{sec:spectroscopy}

We derived stellar parameters (T$_{\rm eff}$, log $g$, and microturbulent velocities), as well as metallicities (taken to be represented by [Fe/H]) for the studied stars using standard techniques employed in quantitative stellar spectroscopy, which relies on equivalent width (EW) measurements of selected samples of Fe I and Fe II lines. 

\begin{figure}
\plotone{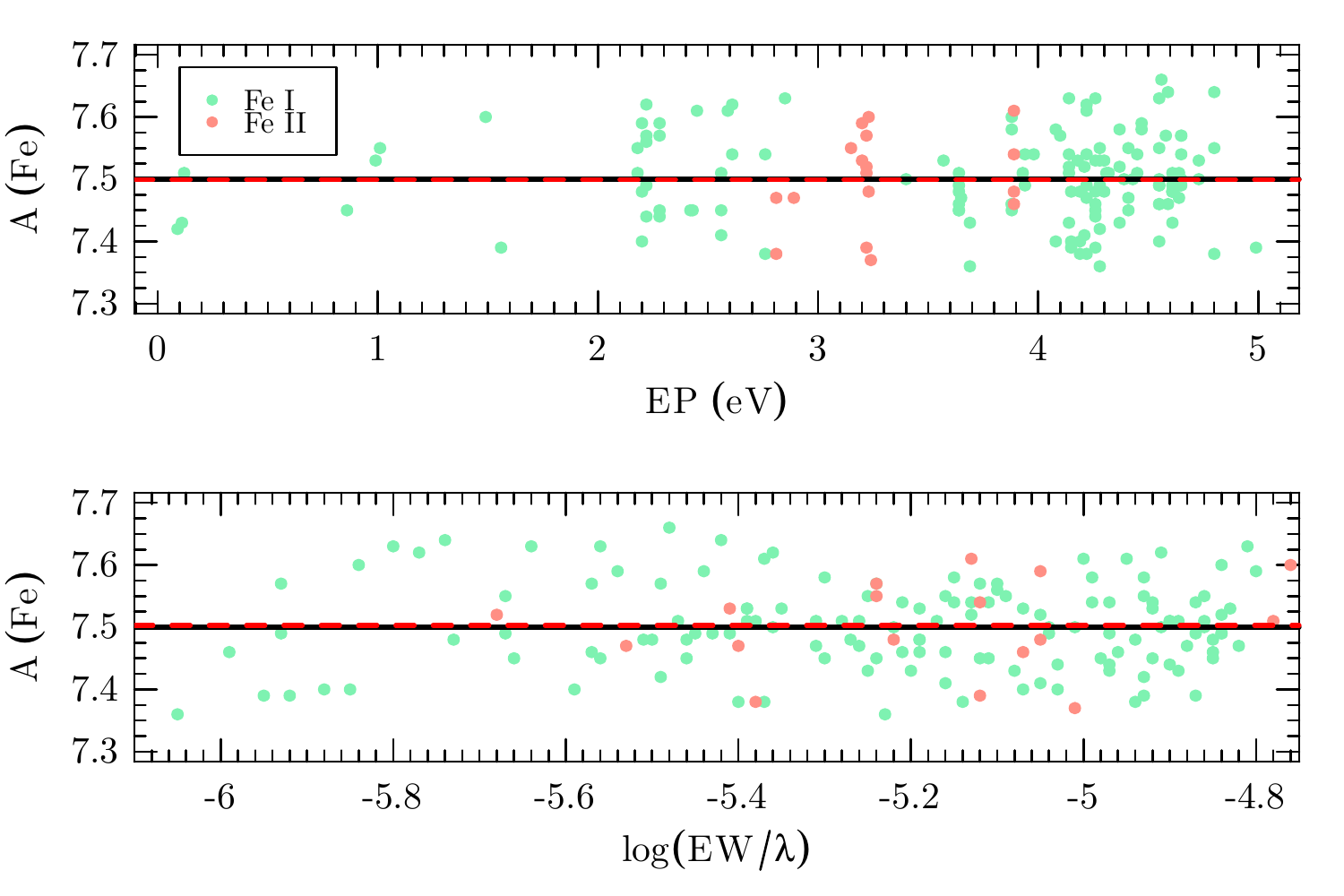}
\caption{An example of the applied methodology to KOI-1, one of the stars in our sample. 
Iron abundances (A(Fe I)) as a function of the excitation potential (EP) of the Fe I and Fe II transitions (top panel) and the reduced equivalent width of the lines (EW/$\lambda$; bottom panel). The atmospheric parameters, iron abundances and microturbulent velocities are obtained once the correlation coefficients of the dashed lines show no dependence with the line parameters.}
\label{fig:method}
\end{figure}

All abundance calculations were done under the assumption of local thermodynamic equilibrium (LTE) using 1D model atmospheres.
The atmospheric parameters were obtained by iterating until the line-by-line Fe I abundances, A(Fe I), exhibit no dependence with the excitation potential of the transitions ($EP$; excitation equilibrium) and, at the same time, the values of A(Fe I) show no trend with the reduced e\-qui\-va\-lent widths (log(EW/$\lambda$)), while finally, the mean Fe I and Fe II abundances reach agreement (ionization equilibrium). These 3 conditions define the stellar T$_{\rm eff}$, log $g$, and microturbulent velocity ($\xi$).
Fig. \ref{fig:method} shows an example of the iterated solution for the effective temperature, log $g$, and microturbulent velocity, as well as the abundance of iron for sample star, KOI-1.


In order to analyze the large number of stars in our sample in an efficient and homogeneous way, we used the automated stellar parameter and metallicity pipeline described in detail in \cite{ghezzi2010b,ghezzi2018}. 
This code uses the updated version of the routine ARES \citep{sousa2015} to measure EWs of Fe I and Fe II lines automatically, the abundance analysis code MOOG 
\citep{sneden1973} to compute the iron abundances, and model atmospheres from the Kurucz ATLAS9 ODFNEW grid \citep{castelli2004}. 
In summary, the code starts with a model atmosphere calculated assuming solar values for T$_{\rm eff}$, log $g$, and metallicity and then iterates until obtaining a final adjusted value for the spectroscopic parameters of each star.

The adopted line list in this study was taken from \cite{ghezzi2018} and consists of 158 Fe I and 18 Fe II isolated and unblended lines. The log $gf$ values of the Fe I and II lines were obtained in \cite{ghezzi2018} from an inverse solar analysis using 
a Kurucz ATLAS9 ODFNEW model atmosphere for the Sun (T$_{\rm eff}$ = 5777 K, log $g$ = 4.44, [Fe/H] = 0.00 and $\xi$ = 1.00 km s$^{-1}$) and   
an adopted solar abundance (A(Fe$_{\odot}$) = 7.50) from \cite{asplund2009}.  

\begin{figure}
\epsscale{0.75}
\plotone{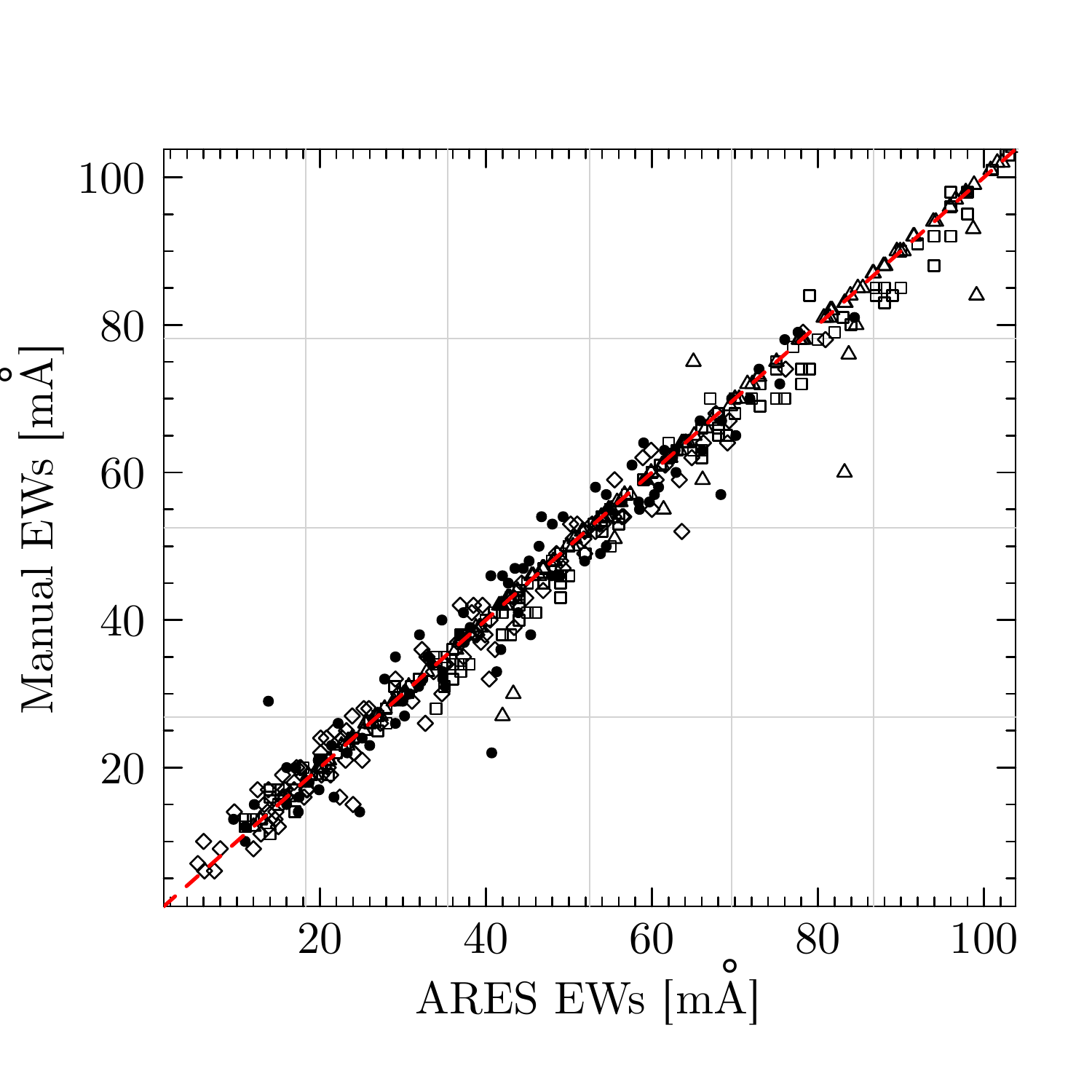}
\caption{Comparison between automatic (using ARES code) and manual equivalent width measurements (using $splot$ task from IRAF)  for a sample of 540 Fe I and Fe II lines in four target stars: KOI-64 (open squares), KOI-268 (filled circles), KOI-280 (open diamonds) and KOI-5782 (open triangles). The red dashed line represents perfect agreement.}
\label{fig:ares}
\end{figure}

As a consistency check, we also measured manually the equivalent widths of a total of 540 Fe I and Fe II in four sample stars (KOI's 64, 268, 280 and 5782 with a mean S/N of 70 in their spectra) using the IRAF package $splot$. A comparison of our manual EW measurements with the automatic ones using the ARES code is presented in Figure \ref{fig:ares}. 
Despite the fact that there were some lines with discrepant equivalent width measurements (these are not an issue in the final solution because the pipeline performs two rounds of sigma-clipping to remove lines with abundances that are too discrepant from the average values) the equivalent widths compared well, showing, on average, a small offset of 1.25 m\AA{} (in the sense of ARES EW being larger than ours) and a RMS scatter of 3.18 m\AA{}.


Table \ref{t:parameter} presents the resulting effective temperatures, surface gravities, microturbulent velocities, and stellar radii (these will be discussed in Section \ref{sec:radii}), and their respective uncertainties, for all stars in our sample. 
The metallicities obtained for the CKS sample will be presented and discussed in a forthcoming paper (Martinez et al. 2019 - Paper II).

\begin{deluxetable*}{lcccchhcccc}[b!]
\tablecaption{Stellar Parameters and Radii\label{t:parameter}}
\tablecolumns{11}
\tablenum{1}
\tablewidth{0pt}
\tablehead{
\colhead{$KOI$} & \colhead{$T_{\rm eff}$} & \colhead{$\delta T_{\rm eff}$} & 
\colhead{log $g$} & \colhead{$\delta$log $g$} & \nocolhead{[Fe/H]} & \nocolhead{$\sigma$[Fe/H]} & \colhead{$\xi$} & 
\colhead{$\delta\xi$} & \colhead{$R_{\star}$} & \colhead{$\delta R_{\star}$} \\
\colhead{No.} & \colhead{(K)} & \colhead{(K)} & \colhead{(dex)} & \colhead{(dex)} & \nocolhead{} & \nocolhead{} & \colhead{(km\,s$^{-1}$)} & \colhead{(km\,s$^{-1}$)} & \colhead{($R_{\odot}$)} & \colhead{($R_{\odot}$)}
}
\startdata
 K00245 & 5410 &  11 & 4.55 & 0.02 & -0.32 & 0.01 & 0.83 & 0.04 & 0.78 & 0.01 \\ 
 K01925 & 5418 &  18 & 4.48 & 0.06 & 0.07 & 0.01 & 0.82 & 0.04 & 0.90 & 0.02 \\ 
 K01612 & 6167 &  35 & 4.32 & 0.09 & -0.19 & 0.02 & 1.43 & 0.06 & 1.24 & 0.03 \\ 
 K00069 & 5646 &  21 & 4.50 & 0.04 & -0.14 & 0.01 & 0.91 & 0.03 & 0.94 & 0.02 \\ 
 K03167 & 5467 &  29 & 4.48 & 0.07 & -0.05 & 0.02 & 1.03 & 0.06 & 0.88 & 0.02 \\ 
 K00082 & 4849 &  49 & 4.51 & 0.08 & 0.03 & 0.02 & 0.62 & 0.12 & 0.73 & 0.01 \\ 
 K00975 & 6227 &  42 & 4.00 & 0.03 & -0.10 & 0.01 & 1.50 & 0.06 & 1.95 & 0.08 \\ 
 K02687 & 5800 &  22 & 4.55 & 0.06 & 0.04 & 0.01 & 1.02 & 0.03 & 0.94 & 0.02 \\ 
 K01924 & 5973 &  28 & 3.86 & 0.06 & 0.27 & 0.01 & 1.53 & 0.03 & 2.57 & 0.14 \\ 
\nodata  & \nodata  & \nodata & \nodata  & \nodata  & \nodata  & \nodata  & \nodata & \nodata & \nodata &  \nodata\\ 
\enddata
\tablecomments{This table is published in its entirety in the machine readable format. A portion is shown here for guidance regarding its form and content.}
\end{deluxetable*}

\subsection{Stellar and Planetary Radii} \label{sec:radii}

Precise stellar radii result in precise planetary radii - a crucial parameter necessary to unveil planetary composition and to ultimately define the transition from rocky to gaseous planets.

To calculate the stellar radii (R$_{\star}$) we used the Stefan-Boltzmann law, which depends on the Stefan-Boltzmann constant ($\sigma_{sb}$), stellar effective temperature (T$_{\rm eff}$) and luminosity (L$_{\star}$),

\begin{displaymath}
     R_{\star} = (\frac{L_{\star}}{4\pi\sigma_{sb}T_{\rm eff}^{4}})^{1/2}  
\end{displaymath}

where L$_{\star}$

\begin{displaymath}
    L_{\star} = L_{0} 10^{-0.4M_{bol}}
\end{displaymath}

L$_{0}$ is the zero point of bolometric magnitude scale \citep{mamajek2015} and M$_{bol}$ the bolometric magnitude related with photometric apparent magnitude (m$_{k}$), extinction (A$_{k}$) and bolometric correction (BC) in the same band, and distance modulus ($\mu$) via 

\begin{displaymath}
      M_{bol} = m_{k} - A_{k} - \mu + BC_{k}
\end{displaymath}

For each target star we used the distances estimated by \cite{bailer2018}, when available. \cite{bailer2018} adopted a bayesian approach and geometric priors to obtain the distances from the $Gaia$ DR2 parallaxes and considered the systematic parallax offsets determined from $Gaia$'s observations of quasars \citep{lindegren2018,zinn2018}.
We used the 2MASS K-band, combined with the reddening E(B-V) derived from the 3D dust map of \cite{green2018} and transformed into A$_{k}$ extinction using the relations in \cite{bilir2008}; and the $isoclassify$ package \citep{huber2017} in its ``direct mode'' using T$_{\rm eff}$, log $g$, [Fe/H] and A$_{v}$ extinction as inputs to interpolate bolometric corrections from MIST grids \citep{choi2016} and calculate absolute magnitudes and stellar luminosities.

In order to calculate the stellar radii, we combined all these parameters with our derived T$_{\rm eff}$; the results are presented in Table \ref{t:parameter}.

Finally, we used the stellar radii to determine planet radii using the values of transit depth, $\Delta$F, which are the fraction of stellar flux lost at the minimum of the planetary transit, cataloged in \cite{thompson2018}; and the equation from \cite{seager2003}:

\begin{equation}
   R_{pl} = 109.1979 \times (\Delta F \times 10^{-6})^{1/2} \times R_{\star}
\end{equation}

To assure a higher level of reliability in the computed planetary radii, we only included in our sample those planets whose host stars have $\leq$10$\%$ error in their derived stellar radius (see Figure \ref{fig:hr}). 
In addition, we removed KOI's that have been classified as ``false positives'' (adopting the same dispositions as in \citealt{thompson2018}).
The derived radii for the final planetary sample in this study are presented in Table \ref{t:pl_rad}.

\begin{deluxetable*}{lcc}
\tablecaption{Planetary Radii \label{t:pl_rad}}
\tablecolumns{3}
\tablenum{2}
\tablewidth{0pt}
\tablehead{
\colhead{Planet} &
\colhead{R$_{pl}$} &
\colhead{$\delta$ R$_{pl}$} \\
\nocolhead{} & \colhead{(R$_{\oplus}$)} & \colhead{(R$_{\oplus}$)}
}
 \startdata
 K00001.01 & 13.62 & 0.30 \\ 
 K00002.01 & 17.93 & 1.07 \\ 
 K00007.01 & 4.49 & 0.13 \\ 
 K00010.01 & 16.62 & 1.14 \\ 
 K00017.01 & 14.31 & 0.35 \\ 
 K00018.01 & 16.78 & 0.80 \\ 
 K00020.01 & 21.19 & 0.63 \\ 
 K00022.01 & 14.06 & 0.34 \\ 
 K00041.01 & 2.48 & 0.07 \\ 
 K00041.02 & 1.38 & 0.04 \\ 
 \nodata & \nodata & \nodata \\
\enddata
\tablecomments{This table is published in its entirety in the machine readable format. A portion is shown here for guidance regarding its form and content.}
\end{deluxetable*}

\subsection{Uncertainties in the Derived Parameters} \label{sec:errors}

The internal errors in the derived effective temperatures and microturbulent ve\-lo\-ci\-ties were calculated by changing these parameters until the slopes of both the A(Fe I) versus $EP$, and A(Fe I) versus log(EW/$\lambda$), respectively, reach the same values as the errors in the slopes from the converged solution.
The uncertainties in log $g$ were estimated by varying this parameter until the Fe I and Fe II mean abundances differed by one standard deviation of the mean of the A(Fe I).

\begin{deluxetable*}{cc}
\tablecaption{Error Budget \label{t:error}}
\tablecolumns{2}
\tablenum{3}
\tablewidth{0pt}
\tablehead{
\colhead{Parameter} &
\colhead{Median Uncertainty} 
}
 \startdata
m$_{k}$       & 0.022 mag \\
A$_{k}$       & 0.009 mag \\
BC$_{k}$      & 0.03 mag  \\
$\mu$         & 0.006 mag \\
T$_{\rm eff}$ & 40 K \\
R$_{star}$    &  2.8$\%$  \\
$\Delta F$    &  4$\%$    \\
R$_{pl}$      &  3.7$\%$ \\
\enddata
\end{deluxetable*}

In order to estimate the total error budget in the derived stellar and planetary radii for our sample, we consider the individual contributions of the errors in each one of the parameters used in the computation of the radii. (See also the discussion in \citealt{fulton2018}). 

The error in the $K$-band stellar magnitudes (m$_{k}$) contributes with 1$\%$ to the stellar radius error if we consider 0.022 mag to be the median error in m$_{k}$ from 2MASS \citep{skrutskie2006} for our target stars.
The contribution due to errors in the extinction A$_{k}$ is even smaller, given that the A$_{k}$ values for the target stars are quite small, ranging between 0.001 to 0.052 mag, with a median A$_{k}$ of 0.009 mag. If we were to completely neglect extinction, this would result in an error of 0.6$\%$ in the stellar radii (using the median A$_{k}$ in the estimate).   
The errors in the $K$-band bolometric corrections (BC$_{k}$) are mainly dominated by uncertainties in the effective temperatures.  
We estimate an error in BC$_{k}$ in the same manner as in \cite{fulton2018}, by changing the effective temperature of a solar-type star by the median error in our T$_{\rm eff}$'s (40 K) and investigating the corresponding change in BC$_{k}$.
(We also investigated the effect of the errors in log $g$ and metallicities, but these were found to be negligible). 
Taking the error obtained in BC$_{k}$ for a test solar type star as typical, we estimate a change of 0.03 mag in BC$_{k}$ (\citealt{huber2017} also estimate the error in BC$_{k}$ to be 0.03 mag) and an error of 0.3 $\%$ for the stellar radius.
 
The distances and the respective errors for the target stars were taken directly from \cite{bailer2018}; and correspond to a median error of 0.006 mag for the ``distance modulus'' ($\mu$) and 0.08$\%$ error in R$_{\star}$.
The internal precision (median error) in the effective temperatures in this study is 40 K, corresponding to 2$\%$ in the R$_{\star}$ error. 

Combining all of the errors in the parameters discussed above in quadrature and propagating the errors, we obtain a median internal uncertainty in our derived stellar radii distribution of $\sim$ 2.8$\%$. 
This uncertainty in the stellar radii has a direct impact, along with the transit depth ($\Delta$F) errors, on the determination of the planetary radii errors. 
We adopted the transit depth values $\Delta$F and respective errors from \cite{thompson2018}, which, for the planets in our sample, result in 4$\%$ internal precision in $\Delta$F and corresponds to a 3.7$\%$ internal precision for the R$_{pl}$ error budget. A summary of the contributions to the error budgets in the R$_{\star}$ and R$_{pl}$ determinations are presented in Table \ref{t:error}.

\section{Results}

\begin{figure*}
\epsscale{1.1}
\plotone{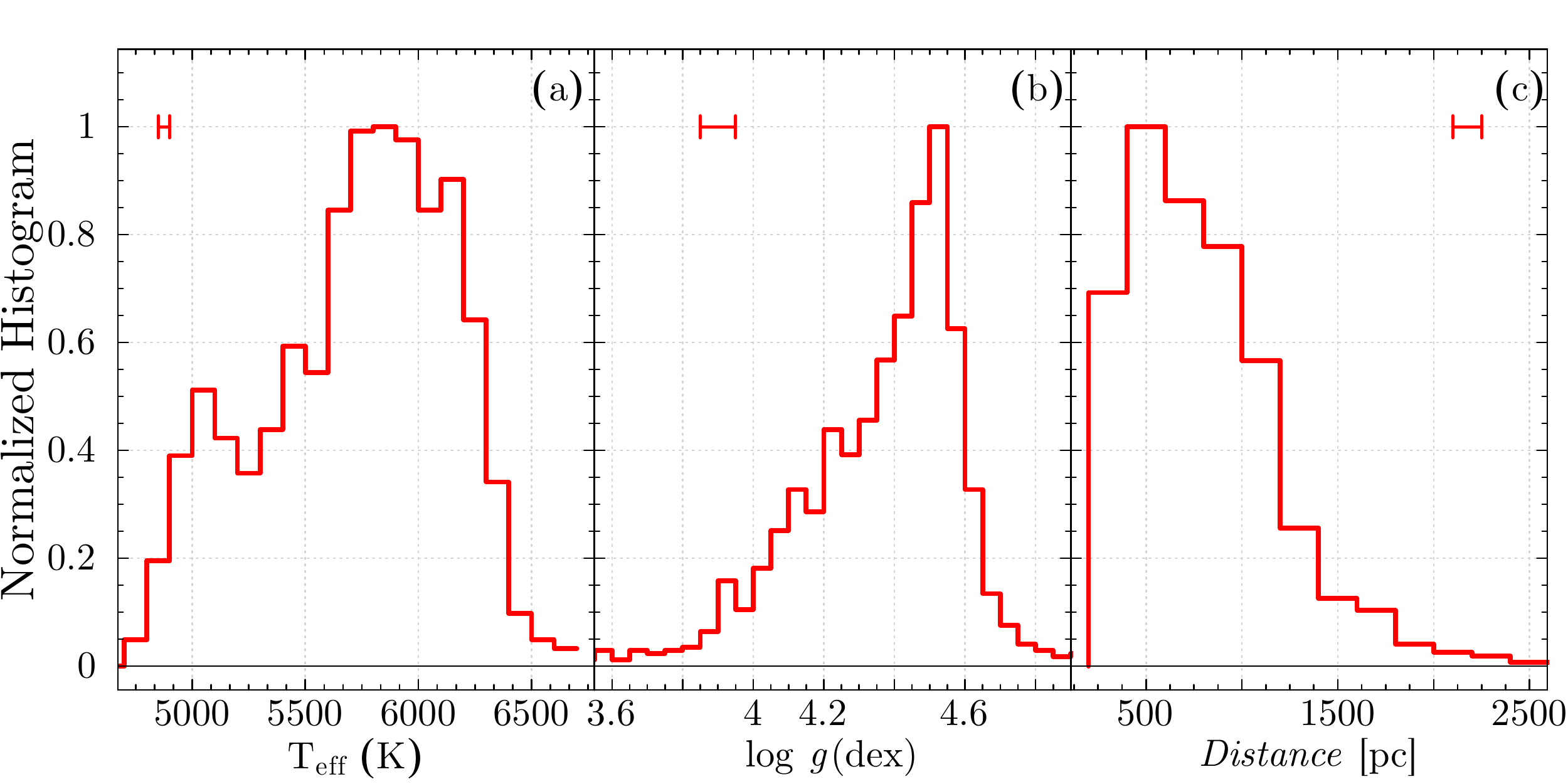}
\caption{Effective temperature (panel a) and surface gravity (panel b) distributions for the sample stars. Panel (c) has the distribution of the stellar distances for the sample; with distances obtained from $Gaia$ DR2 parallaxes and taken from  \cite{bailer2018}. The median uncertainties in the parameters are plotted in the upper corner of each panel.}
\label{fig:hist}
\end{figure*}

The atmospheric parameter distributions for the studied sample are shown in the different histograms of Figure \ref{fig:hist}; most of the target stars have effective temperatures roughly between 4800 - 6500 K, having a peak around the solar effective temperature (T$_{\rm eff}$ = 5777 K)  and a smaller peak corresponding to cooler stars at T$_{\rm eff}$ $\sim$ 5000 K. The log $g$ distribution for the studied sample corresponds mostly to unevolved stars, with the distribution having a peak between roughly log $g$ = 4.3 -- 4.5 dex, but it also contains a tail with more evolved stars having log $g$ $<$ $\sim$4.2.  
Most of the target stars (about 80$\%$) are from the solar neighborhood having distances \citep[from][]{bailer2018} generally within $\sim$1 Kpc of the Sun (Fig. \ref{fig:hist}, panel c).

\begin{figure}
\epsscale{0.85}
\plotone{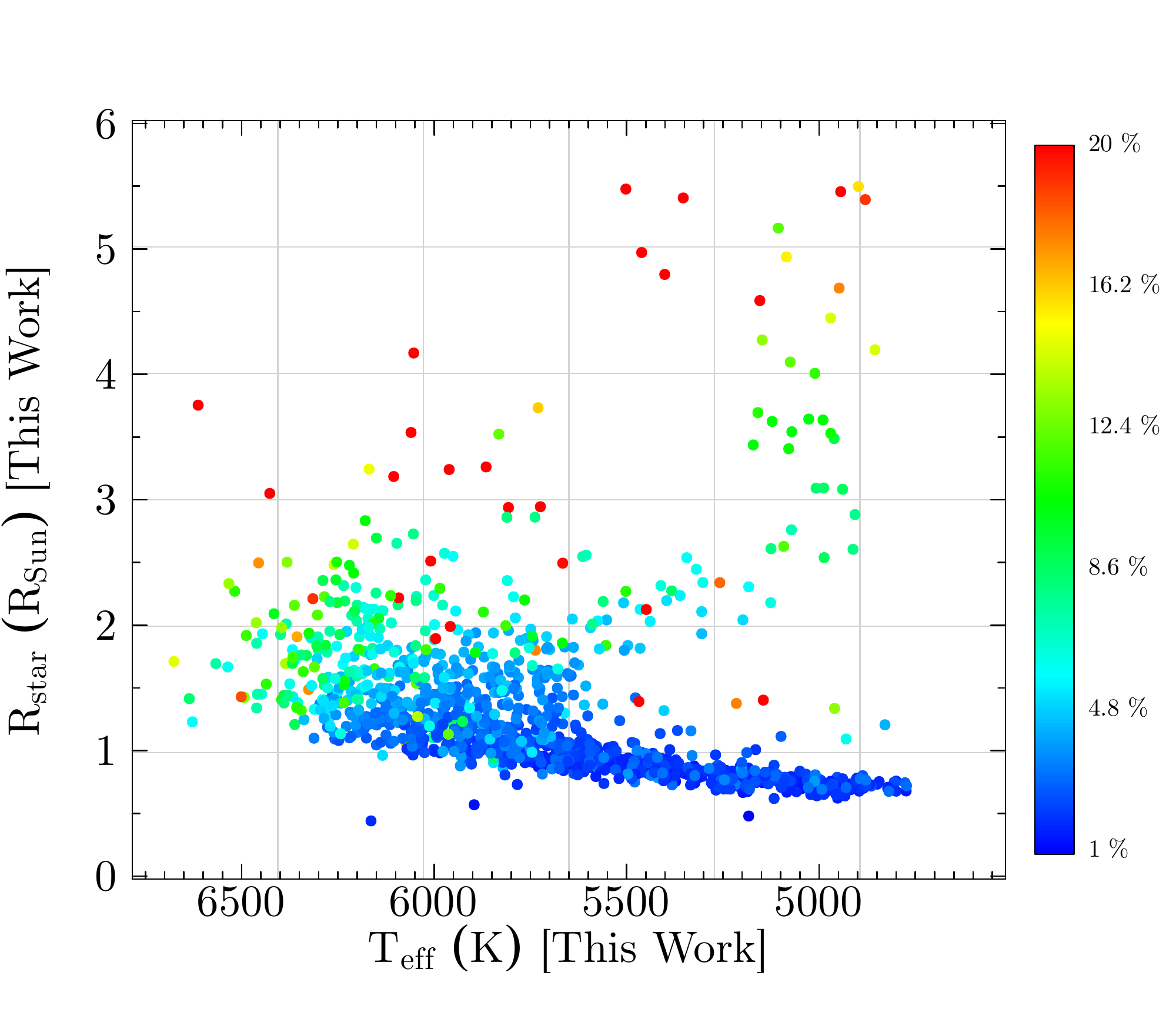}
\caption{The stellar radii versus the effective temperatures for the target stars. The stellar radii are based on $Gaia$ DR2 parallaxes and are shown color-coded by the stellar radii errors: from $\sim$1$\%$ error in blue to 20$\%$ error or larger (up to $\sim$200$\%$) in red.}
\label{fig:hr}
\end{figure}

As discussed in the previous section, the stellar radii in this study were calculated using distances based on $Gaia$ DR2 parallaxes.
In Figure \ref{fig:hr} we present a HR diagram showing the derived effective temperatures and stellar radii, color coded by the errors in the stellar radii. The median internal uncertainty in the radii distribution in this study is $\sim$ 2.8$\%$ (Section \ref{sec:errors}) and most target stars follow the expected sequences in the HR diagram, showing a densely populated main sequence and the presence of some evolved stars (as expected from the T$_{\rm eff}$ and log $g$ distributions obtained for the targets and shown in Fig. \ref{fig:hist}). There are, however, clear outliers that occupy unexpected loci in the diagram corresponding to at least 10$\%$ error in the stellar radii (represented by green, yellow and red points in Figure \ref{fig:hr}).


\subsection{Previous Results from the Literature}

\subsubsection{Stellar Parameters Comparisons for the CKS Sample}

As previously mentioned, the study by \cite{petigura2017} was the first to present a spectroscopic analysis of the CKS sample.
That study derived spectroscopic parameters using two different spectral synthesis techniques in LTE: SpecMatch and SME@XSEDE. This methodology is very distinct from the one used in our study.
SpecMatch, was specifically designed for the CKS project. 
In summary, SpecMatch fits model spectra (computed with Kurucz models by \citealt{coelho2005}) in\-di\-vi\-dua\-lly to five different wavelength segments in the observed spectra, and averages the resulting sets of parameters T$_{\rm eff}$, log $g$, metallicity, and v sin$i$ of each segment. The results from the \cite{brewer2016a} catalog were used to calibrate the SpecMatch results. 

SME@XSEDE is an automated version of the spectral synthesis code Spectroscopy Made Easy \citep[SME,][]{valenti1996}. It uses a line list with atomic parameters taken from the VALD database to interpolate between a grid of plane-parallel MARCS model atmospheres \citep{gustafsson2008} until the optimal solution is found, using a $\chi^{2}$ minimization.
The final results in \cite{petigura2017} were obtained 
by applying linear corrections to the raw SME@XSEDE results to put them on the SpecMatch scale (originally calibrated to the \citealt{brewer2016a} scale), while for those target stars with consistent results between SpecMatch and SME@XSEDE, the authors adopted the mean value from the two methods.

\begin{figure*}  
\gridline{\fig{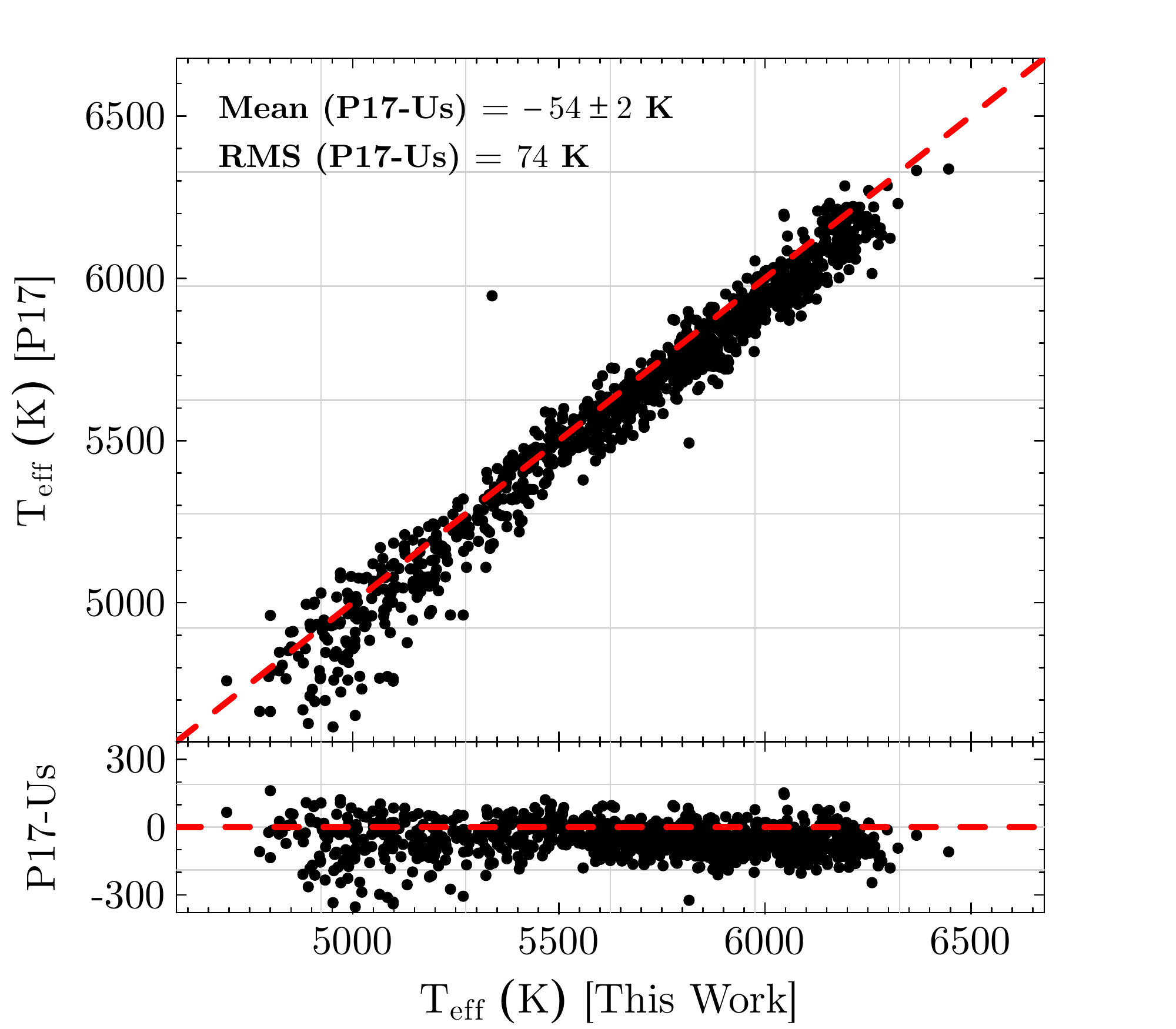}{0.45\textwidth}{}
          \fig{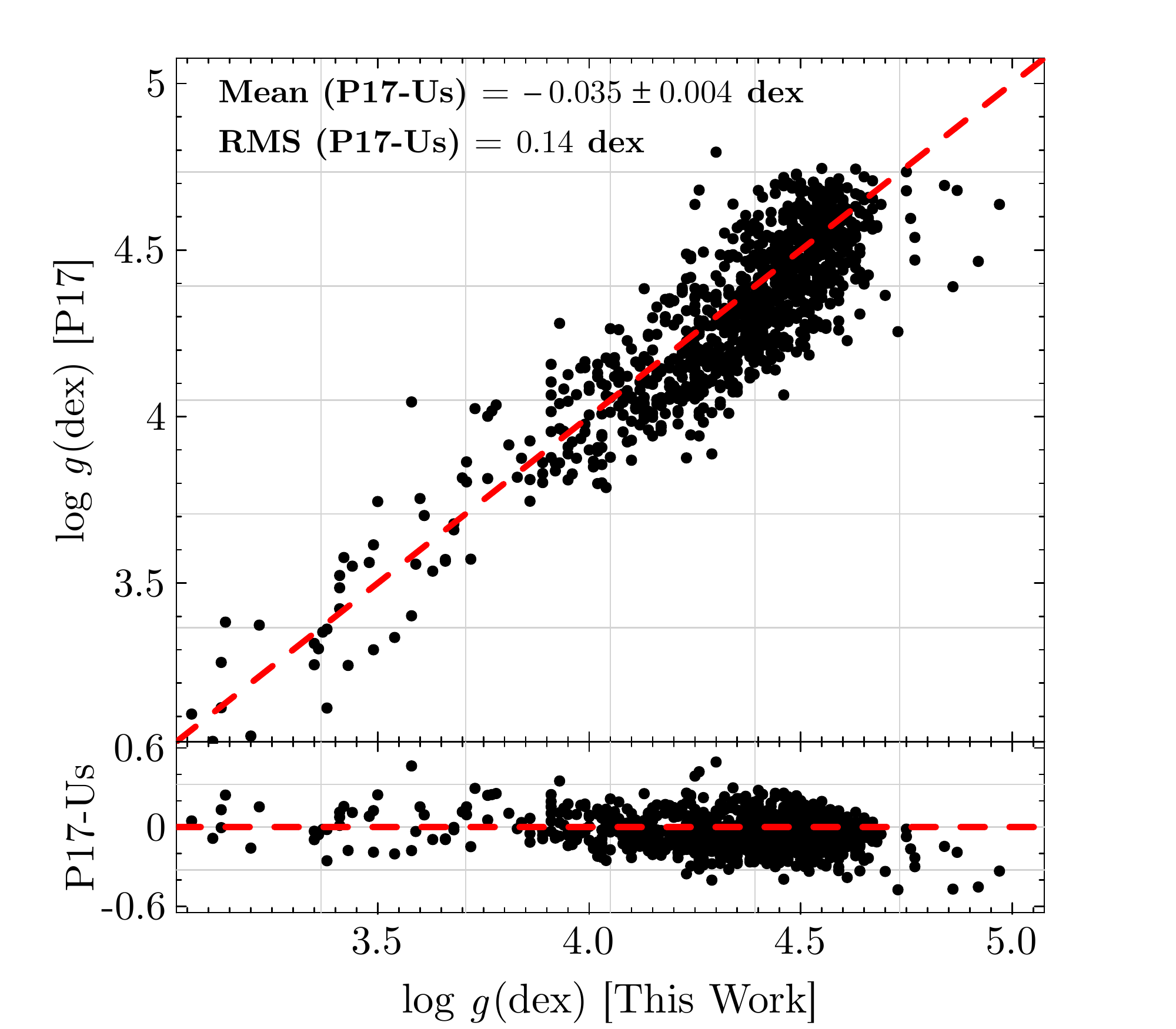}{0.45\textwidth}{}
          }
\gridline{\fig{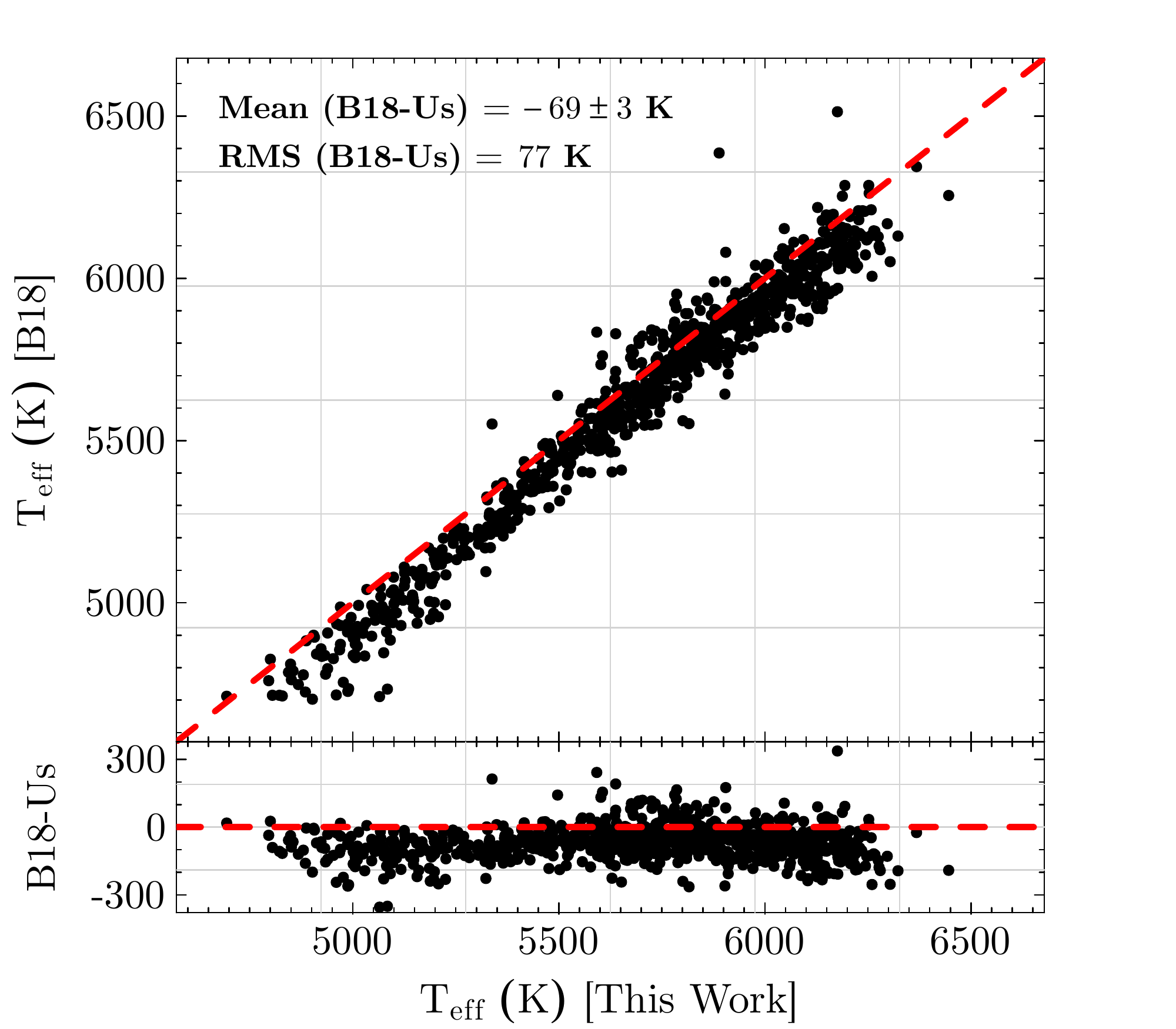}{0.45\textwidth}{(a)}
          \fig{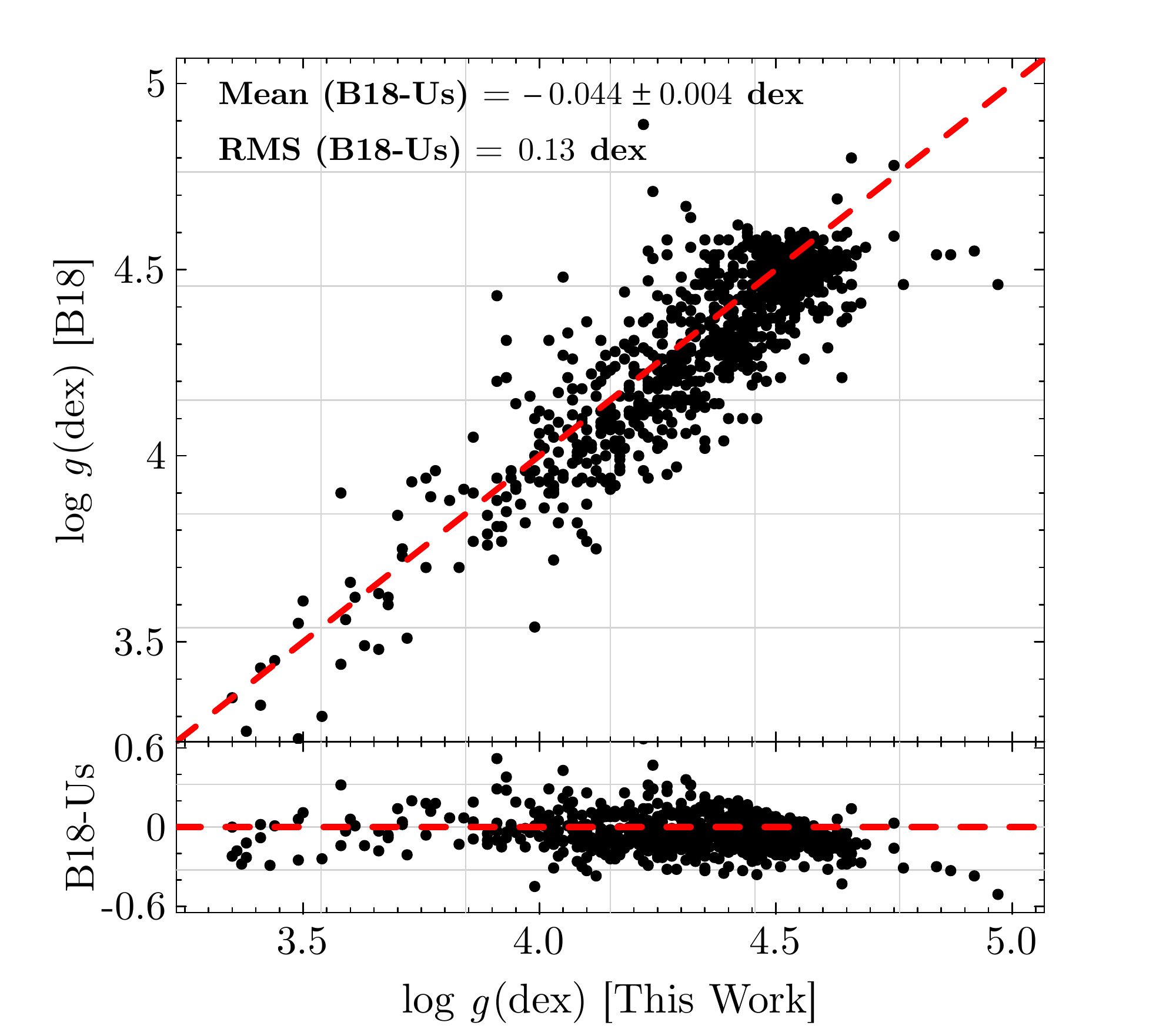}{0.45\textwidth}{(b)}
          } 
\caption{Comparison of the effective temperatures (panels a) and log $g$ values (panels b) derived in this work and in \cite{petigura2017} for 1013 stars in common (top panels) and for 847 stars in common with \cite{brewer2018} (bottom panels). The mean differences between the parameters and the corresponding RMS scatters are indicated in each case. The red dashed lines represent the equality.}
\label{fig:cks}
\end{figure*}

The top panels of Figure \ref{fig:cks} show a comparison of the derived atmospheric pa\-ra\-me\-ters, T$_{\rm eff}$ and log $g$ with results from \cite{petigura2017}.
In general, there is very good agreement (within the uncertainties) with the stellar parameters derived by \cite{petigura2017}, although there is a small systematic difference of about $\sim$60 K in the effective temperatures (our T$_{\rm eff}$ scale being hotter than the \citealt{petigura2017}). 
When considering mean differences for log $g$, there is only a small offset with our derived log $g$ values ($<$\cite{petigura2017} - This Study$>$= -0.035 dex; RMS = 0.14 dex), but it should be noted that twelve stars in our sample have log $g$ $>$ $\sim$4.7, although their log $g$ errors are within the expected uncertainties (mean of the log $g$ errors is 0.12 dex) these results are all systematically higher than \cite{petigura2017}. 

More recently, \cite{brewer2018} also analyzed the CKS dataset. 
They adopted the SME semi-automated spectral synthesis code (also used in \citealt{brewer2016a}) to fit the observed spectra and constrain the stellar parameters. 
The synthesis code uses as input an atomic and molecular line list, a grid of plane-parallel model atmospheres \citep{castelli2004} and 
finds the best solution for global pa\-ra\-me\-ters, such as T$_{\rm eff}$, log $g$, [M/H], or v$_{macro}$.
The authors applied an inverse solar analysis to adjust the oscillator strengths, log $gf$s, of the transitions, and performed a combination of spectral synthesis and asteroseismic techniques to obtain their final results. 

A comparison of our derived stellar parameters with those obtained by \cite{brewer2018} is shown in the bottom panels of Figure \ref{fig:cks} for 847 stars in common. The conclusions are similar to those found with \cite{petigura2017}, which is expected, given that these authors effectively calibrated their results to be on the \cite{brewer2018}, and consequently on the \cite{brewer2016a} scale. For the effective temperatures, the mean difference ($<$BF18 - This study$>$) is -69 $\pm$ 3 K and the RMS = 77 K, again indicating a small systematic offset in the two T$_{\rm eff}$ scales. For log $g$, there is a small systematic difference of -0.044 dex, which is not significant.


\subsubsection{Stellar Parameter Comparisons for Other Samples of Kepler Stars}

\cite{buchhave2012} used multiple observations from high-resolution spectrographs on several telescopes to derive stellar parameters for 152 planet-hosting stars discovered by the $Kepler$ mission.
They derived the stellar parameters, T$_{\rm eff}$, log $g$, [m/H] and v$_{rot}$ using the spectral synthesis code Stellar Parameter Classification (SPC), which uses a library of model atmospheres \citep{kurucz1992} to synthesize the spectrum between 5050 - 5360 \AA{} and measures a cross-correlation function peak that indicates how well the synthetic data reproduce the observed ones. 

\begin{figure*}
\gridline{\fig{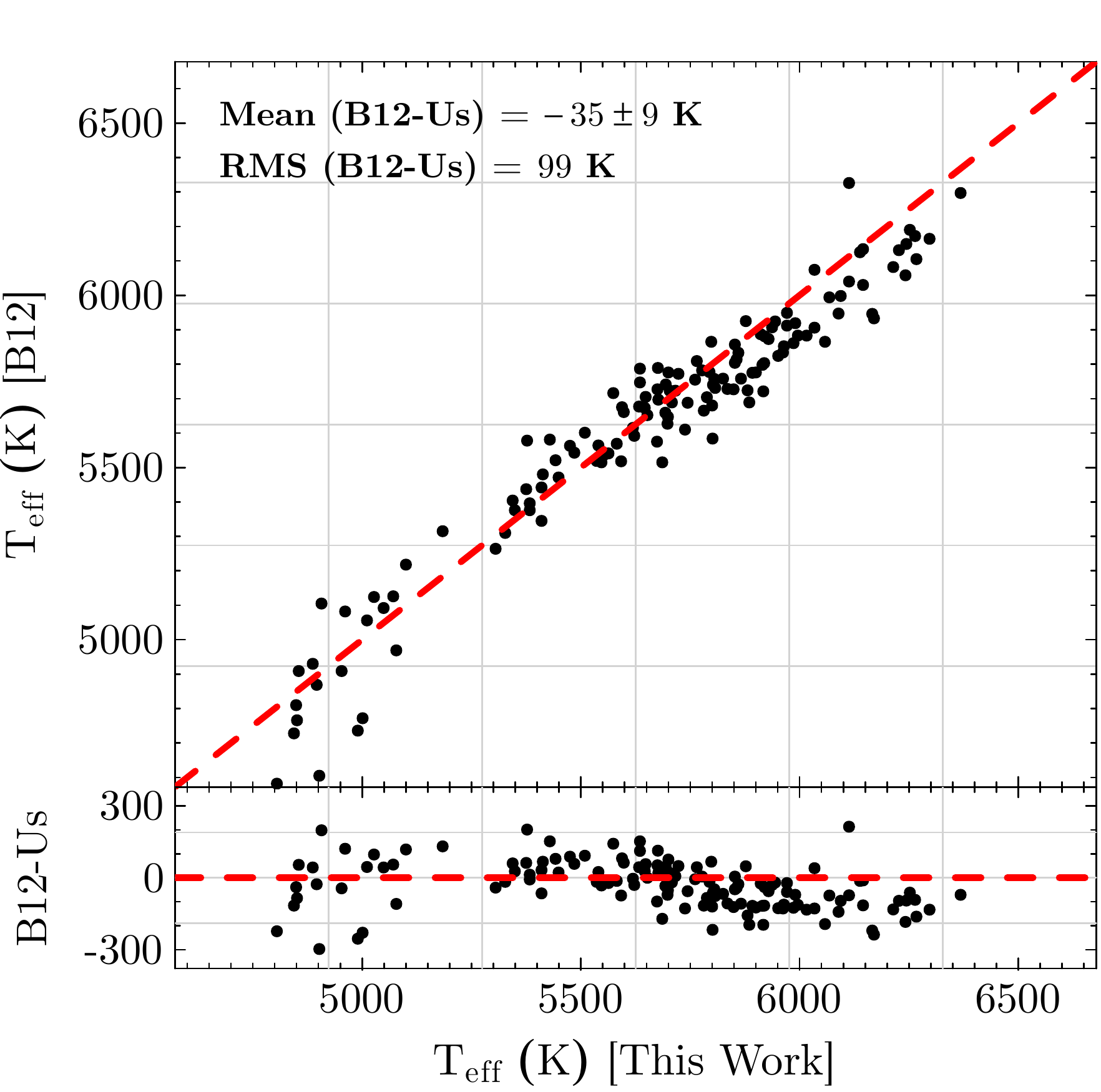}{0.39\textwidth}{}
          \fig{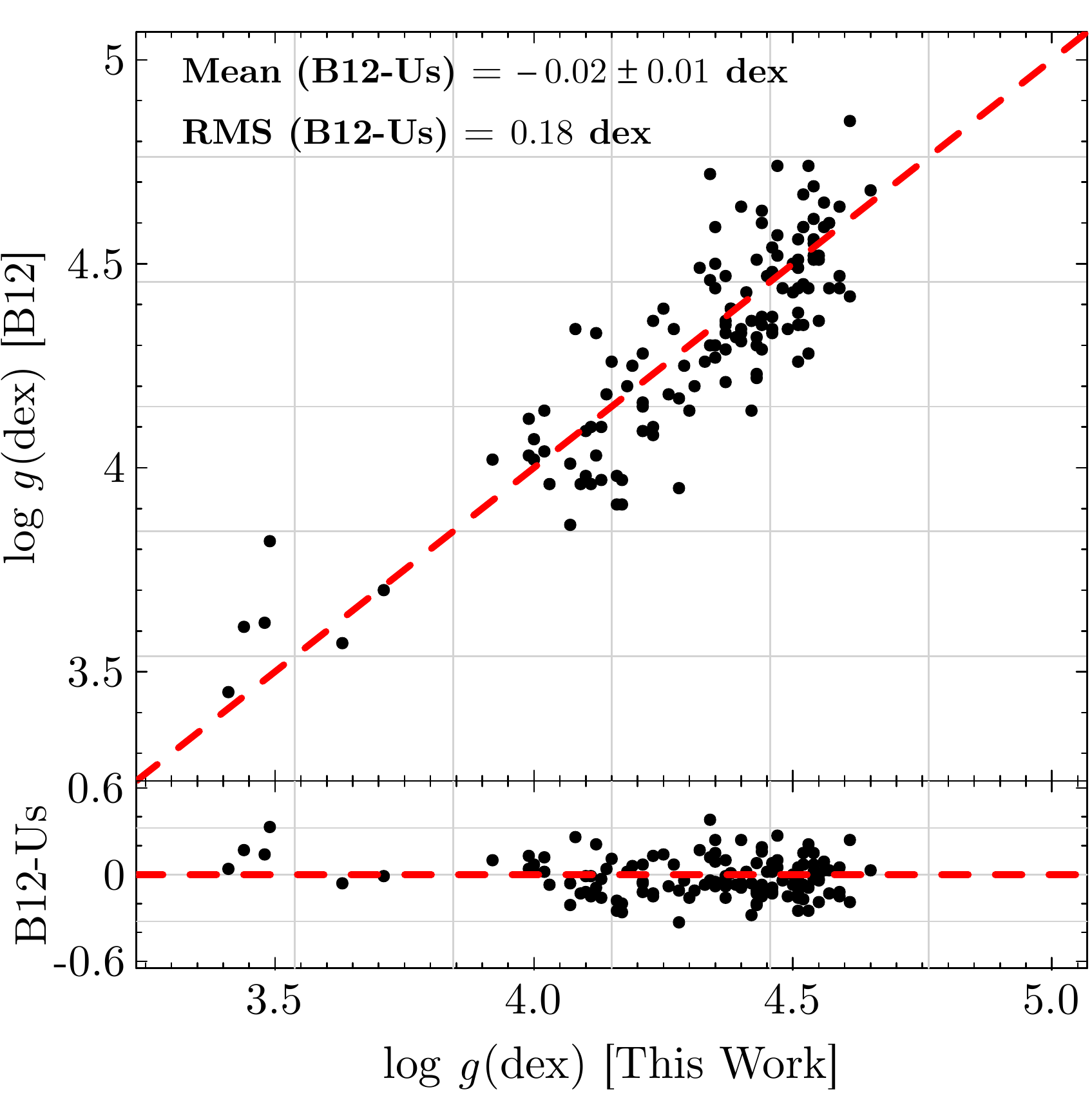}{0.375\textwidth}{}
          }
\gridline{\fig{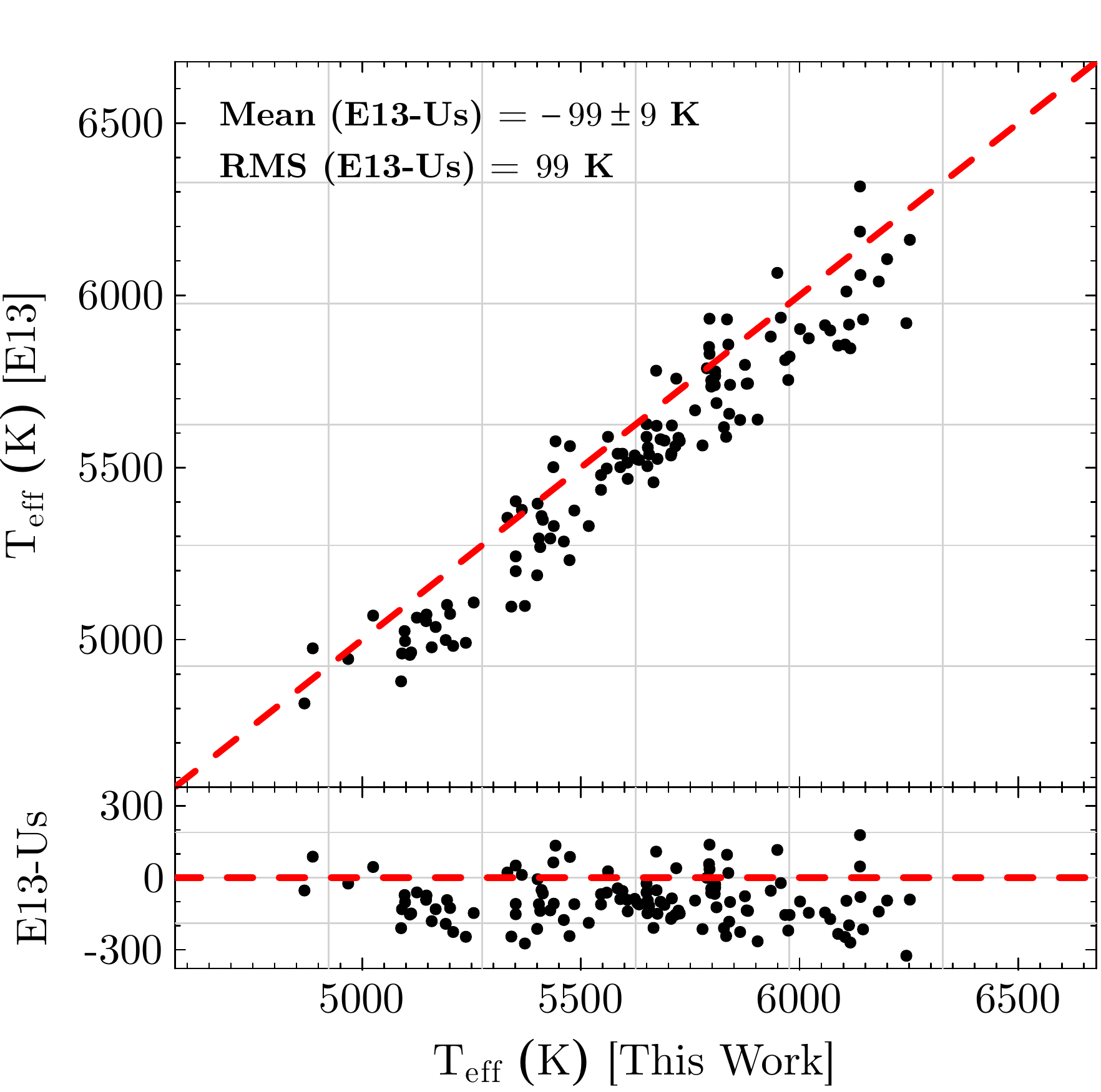}{0.39\textwidth}{}
          \fig{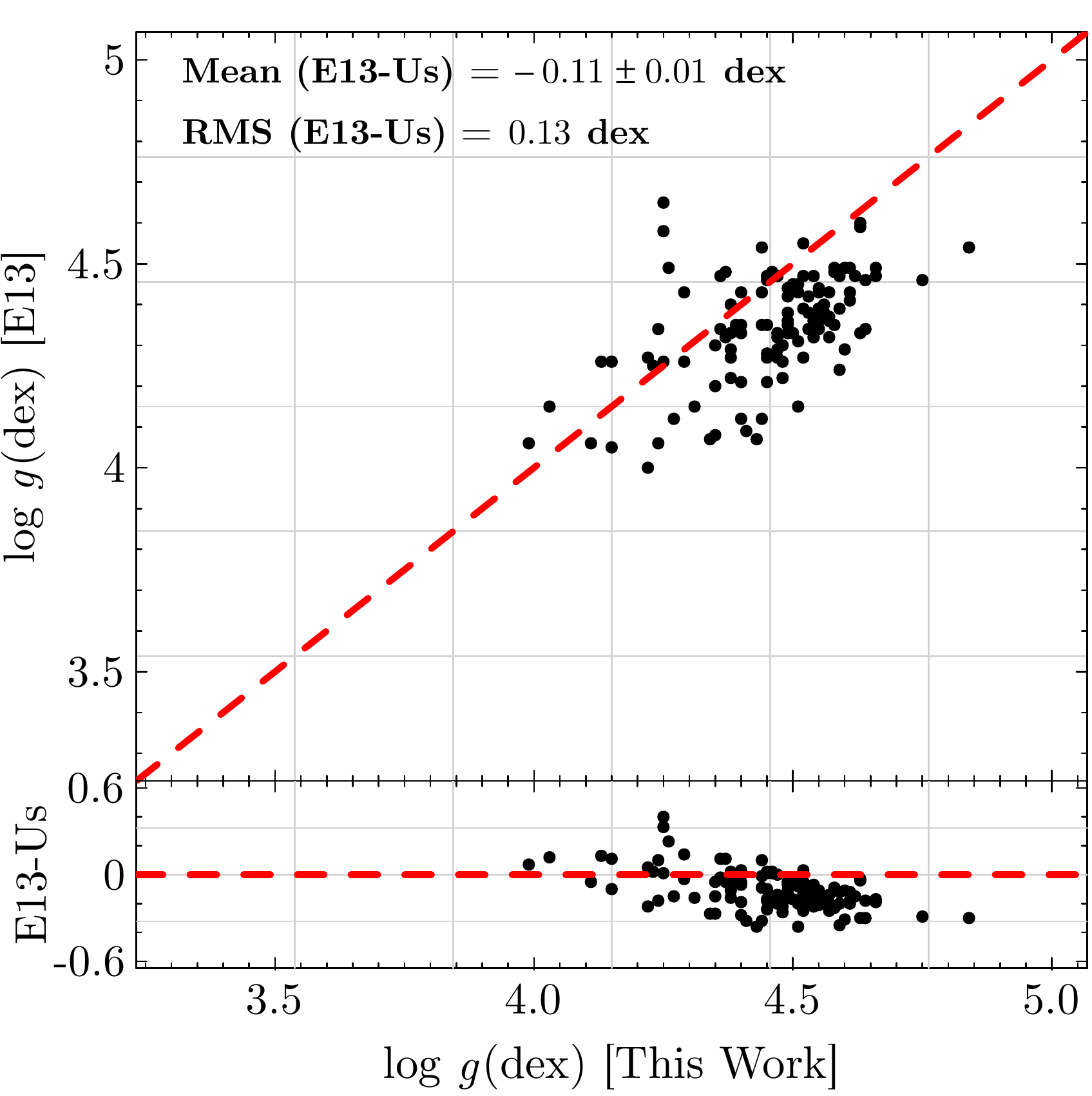}{0.375\textwidth}{}
          }
\gridline{\fig{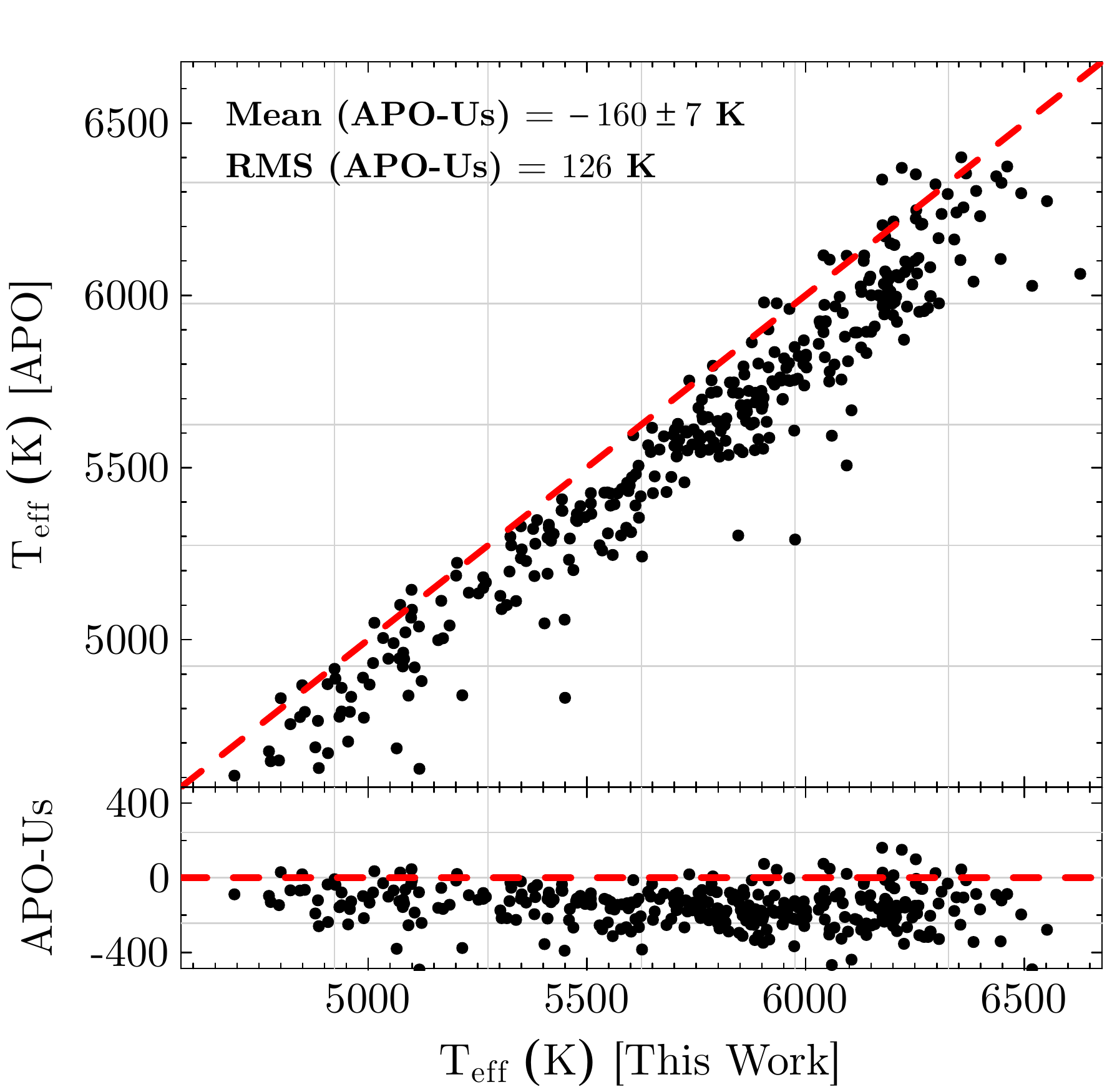}{0.39\textwidth}{(a)}
          \fig{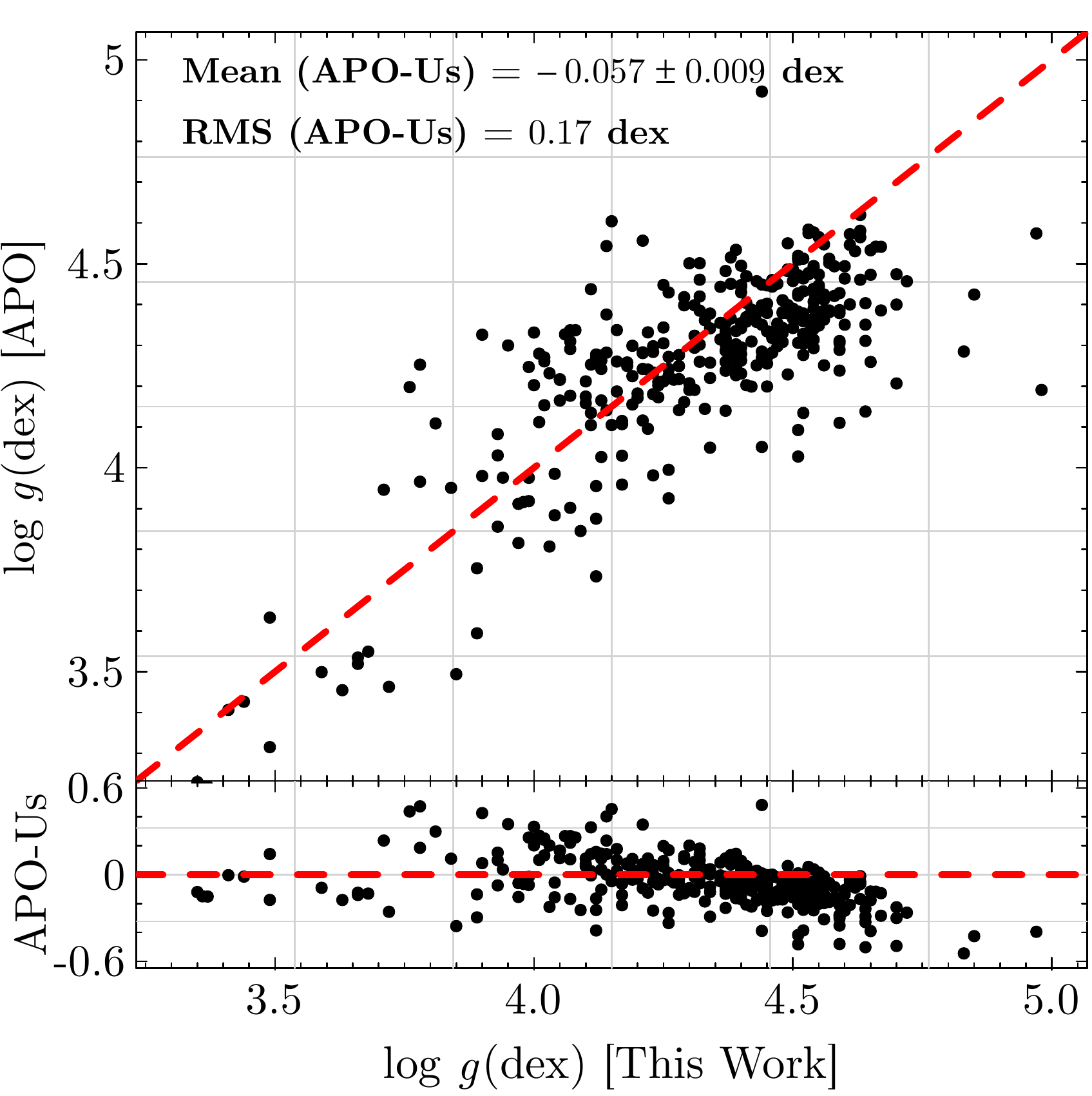}{0.375\textwidth}{(b)}
          }          
\caption{Comparison of the effective temperatures (panels a) and log $g$ values (panels b) derived in this work and in \cite{buchhave2012} for 135 stars in common (top panels), for 119 stars in common with  \cite{everett2013} (middle panels) and for 343 stars in common in the APOGEE DR14 (bottom panels). 
The mean differences between the parameters and the corresponding RMS scatters are indicated in each case. The red dashed lines represent the equality.}
\label{fig:others}
\end{figure*}

The top panels of Figure \ref{fig:others} show that there is a small trend in the comparison of our effective temperature scale with that of \cite{buchhave2012}: at higher temperatures (T$_{\rm eff}$ $>$ 5750 K), our effective temperatures are systematically larger by 82 K $\pm$ 9 K than \cite{buchhave2012}, while for the range between 5200 K and 5750 K our temperatures are systematically lower by 21 K $\pm$ 10 K; for T$_{\rm eff}$'s lower than 5200 K there is no systematic trend but a larger scatter. The agreement with log $g$ is good, with an insignificant systematic difference of -0.02 dex but a higher RMS of $\sim$0.18 dex.


The middle panels of Figure \ref{fig:others} compares our results with those from \cite{everett2013} for a sample of 268 faint candidate exoplanet-hosting stars discovered by the $Kepler$ mission. \cite{everett2013} obtained low-resolution spectra (R = 3,000) u\-sing the RCSpec long-slit spectrograph on the 4-m telescope at Kitt Peak Observatory and determined T$_{\rm eff}$, log $g$ and [Fe/H] by fitting the observed spectra with synthetic ones computed from stellar model atmospheres \citep{castelli2003}.
Effective temperatures from \cite{everett2013} results show a more significant mean offset of $\sim$100 K relative to ours, with our T$_{\rm eff}$ scale being hotter, with a similar value for the RMS scatter. 
It is also noticeable that there is a negative trend 
in the log $g$ differences: for log $g$ $>$ $\sim$4.3 dex our derived log $g$'s are systematically larger than theirs (with a mean difference of 0.14 $\pm$ 0.01 dex), while for log $g$ smaller than $\sim$4.3 dex our results are systematically lower (with a mean difference of 0.05 $\pm$ 0.04 dex). 
The bottom panels of Figure \ref{fig:others} compare results for 343 KOI's in common that were observed with the near-infrared ($\lambda$ 1.5 - 1.7 $\mu$m), high-resolution spectroscopic (R $\sim$ 22,500) Apache Point Observatory Galactic Evolution Experiment \citep[APOGEE -][]{majewski2017}, which is a survey in SDSS-IV.

The stellar parameters shown are part of APOGEE Data Release 14 \citep[DR14;][]{holtzman2018} and were derived automatically using the APOGEE Stellar Parameters and Chemical Abundances pipeline \citep[ASPCAP,][]{garcia2016}, which fits the observed spectra to grids of synthetic spectra using $\chi^{2}$ minimization.
There is a significant systematic offset in effective temperatures, with our T$_{\rm eff}$ scale being hotter than that of APOGEE DR14 by 160 K in the mean, with a RMS scatter of 126 K.
There is also an offset in log $g$ of -0.06 dex (RMS = 0.17 dex), with a negative trend in the mean difference $<$APOGEE - This Study$>$ as function of log $g$, in particular for log $g$ values larger than $\sim$4.0. It is well known, however, that the surface gravities from ASPCAP DR14 have systematic offsets for red giants, as well as for dwarfs \citep{holtzman2018,jonsson2018}.

\subsubsection{Asteroseismic vs. Spectroscopic Surface Gravities}

In this study we use the Fe I and Fe II lines in order to derive surface gravities for the stars, with the log $g$ derivation being done concomitantly with the determinations of effective temperatures, microturbulent velocities, and iron abundances (Section \ref{sec:spectroscopy}). Correlations that exist between these parameters can lead to systematic errors in the derived parameters that can be investigated. In fact, one of the stellar parameters that is typically not very well constrained via spectroscopy is the surface gravity.  Asteroseismology, on the other hand, can provide accurate log $g$'s \citep[to 0.05 dex;][]{pinsonneault2018} that can serve as valuable benchmarks to investigate possible systematic offsets in spectroscopic determinations of log $g$ \citep[][and references therein]{chaplin2014,silva-aguirre2015,huber2017,lundkvist2018}.

\begin{figure}
\epsscale{0.85}
\plotone{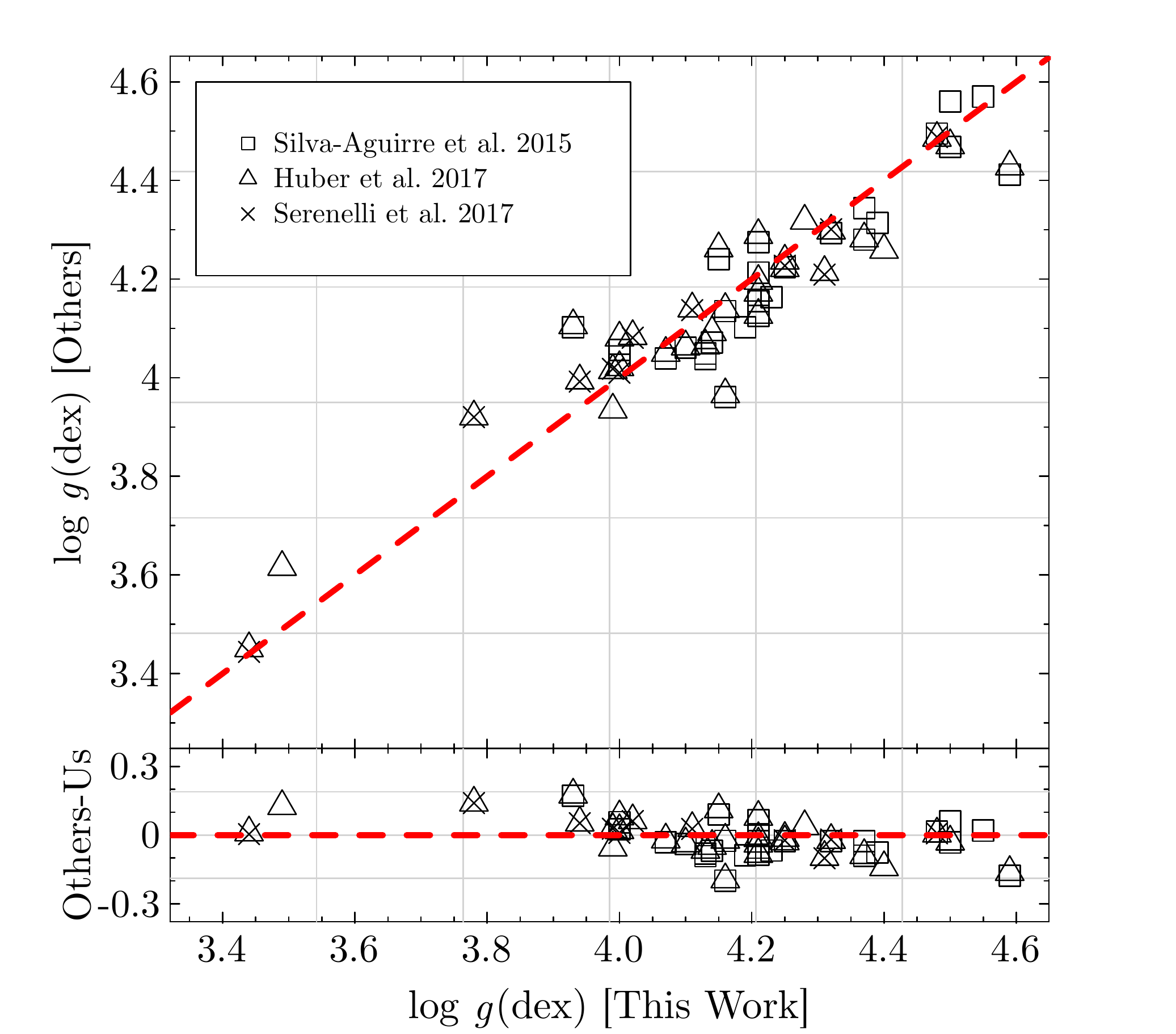}
\caption{Our derived spectroscopic surface gravities in comparison with asteroseismic surface gravities from \cite{silva-aguirre2015}, \cite{huber2017} and \cite{serenelli2017} for 40 stars in common.}
\label{fig:asteros}
\end{figure}

Figure \ref{fig:asteros} shows the comparison between our derived log $g$ values with those determined via asteroseismology. 
The asteroseismic data for 40 sample stars were collected from \cite{silva-aguirre2015}, \cite{huber2017} and \cite{serenelli2017}. 
It should be noted that both T$_{\rm eff}$ and [Fe/H] are essential to constrain the asteroseismic log $g$. 
\cite{huber2017} and \cite{serenelli2017} used the APOGEE results for effective temperatures and metallicities in deriving log $g$, while \cite{silva-aguirre2015} 
used Spectroscopic Made Easy \citep[SME-][]{valenti1996} and Stellar Parameter Classification \citep[SPC-][]{buchhave2012}, two spectral synthesis techniques, to fit optical high-resolution spectra to synthetic ones.

The seismic results presented in Figure \ref{fig:asteros} indicate good agreement between the log $g$ determinations in the 3 different asteroseismic studies (\citealt{silva-aguirre2015}, \citealt{huber2017} and \citealt{serenelli2017}) for the few stars in common. Our derived log $g$ values (based on Fe I and Fe II lines) also compare well with the asteroseismic ones, resulting in an insignificant mean offset in log $g$ ($<$Asteroseismic - This Study$>$) of -0.01 dex and a reasonable RMS of 0.08 dex. Such an agreement is completely within the expected uncertainties in spectroscopic log $g$ determinations.
 
We note that the coverage between log $g$ = 3.4 -- 3.8 is rather limited and the comparison is based only on three stars, two of them with seismic log $g$ values higher than the ones derived here by 0.13 dex. More stars with seismic log $g$'s would be needed in order to reach a firmer conclusion; based on the current data there is no indication of significant offsets in our spectroscopic log $g$ determinations.

\subsubsection{Stellar and Planetary Radii Comparisons for the CKS Sample}

Stellar radii for the CKS sample were previously derived in \cite{johnson2017} by converting the spectroscopic T$_{\rm eff}$, log $g$, and [Fe/H] from \cite{petigura2017} into stellar masses, radii, and ages using the Darmouth Stellar Evolution Program models \citep{dotter2008} interpolated with the $isochrones$ package \citep{morton2015}; \cite{johnson2017} also used their derived stellar radii to determine the planetary radii. \cite{fulton2017} used the planetary radii computed by \cite{johnson2017} to make completeness corrections and calculate the resulting radius distribution. As $Gaia$ DR2 became available, \cite{fulton2018} computed stellar radii using distances from the $Gaia$ DR2 inverted parallaxes and also derived the planetary radii for the CKS sample.

Stellar radii computed taking into account the parallaxes are in principle more precise. \cite{fulton2018} estimated that the errors in their stellar and planetary radii, using distances from inversion of the $Gaia$ parallaxes, are at the level of 2$\%$ and 5$\%$, respectively. Although \cite{fulton2018} obtained a scatter of 13$\%$ in the ratios of stellar radii when compared with those from \cite{johnson2017}, their distribution of planet sizes remained basically the same. 

\begin{figure*}
\plotone{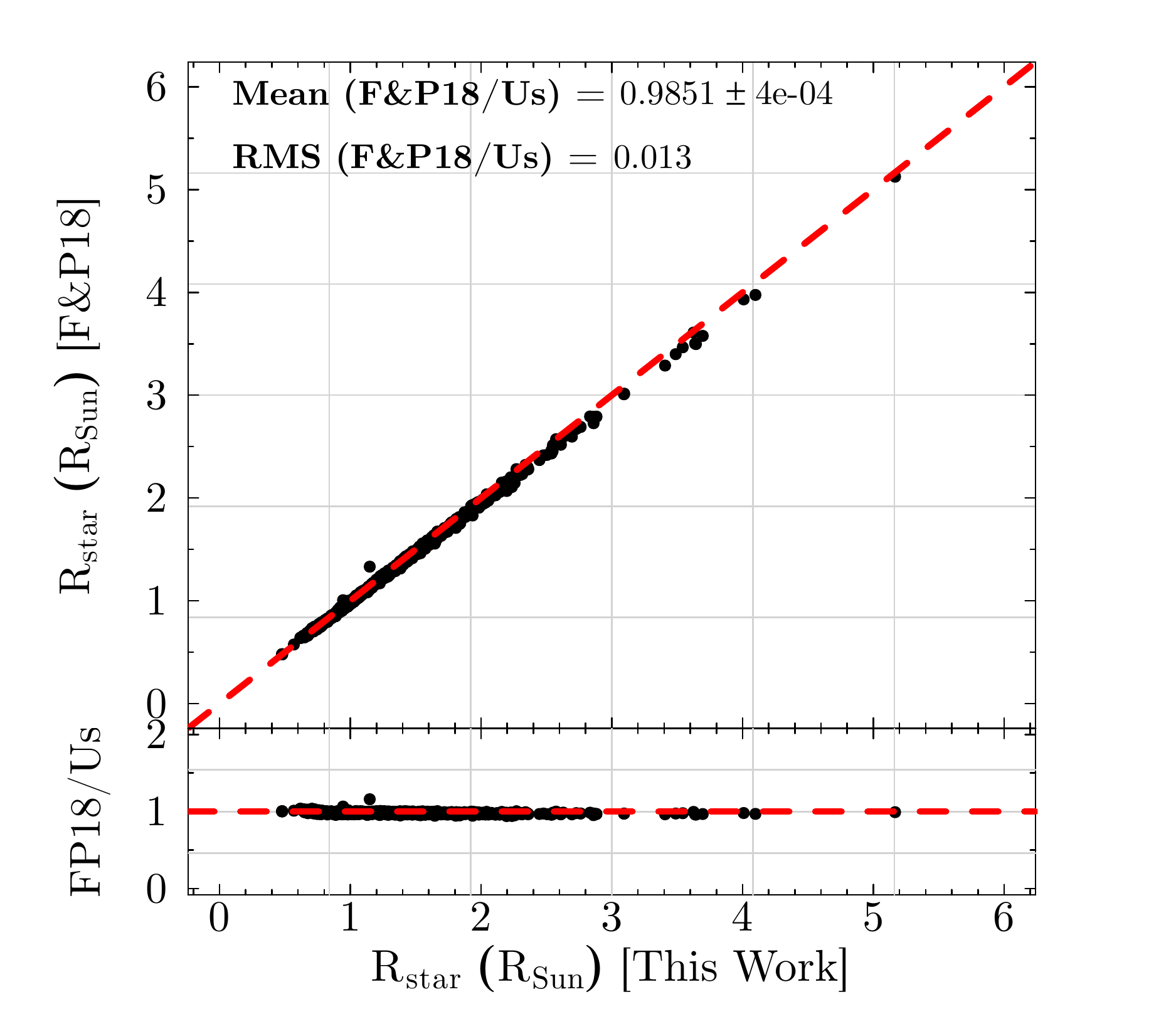}
\caption{A comparison of the stellar radii derived from \cite{fulton2018} with those derived in this work. Our stellar radii are just slightly larger than \cite{fulton2018}. 
The mean ratio between the radii in the studies and the corresponding RMS scatter is shown.
There is a tendency to find larger differences for R$_{\star}$ $>$ 2.5 R$_{\odot}$.}
\label{fig:sr}
\end{figure*}

Figure \ref{fig:sr} shows the comparison between our derived stellar radii with those from \cite{fulton2018}; there is an overall good agreement between the results for the vast majority of the targets; the mean stellar radii ratio between \cite{fulton2018} and this study (``F$\&$P18/Us'') is 0.9851 $\pm$ 0.0004 with a RMS scatter of 0.013.
A closer look at the results in Figure \ref{fig:sr} indicates that the systematic differences are slightly larger for R$_{\star}$ $>$ $\sim$2.5 R$_{\odot}$;
if we compute the mean stellar radii ratio and the RMS for those stars, we obtain: 0.973 $\pm$ 0.002 and 0.010, respectively.

\begin{figure*}
\gridline{\fig{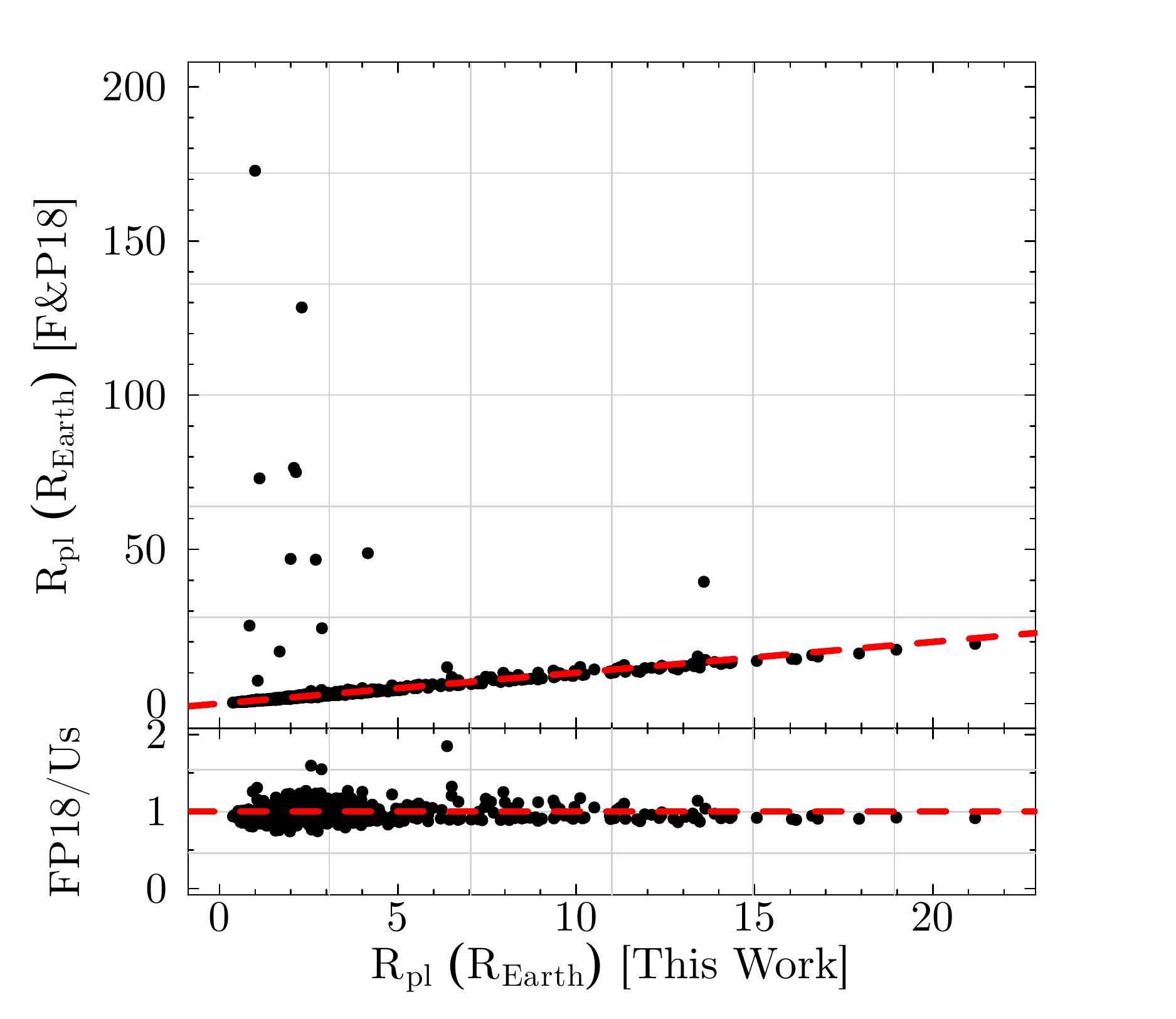}{0.52\textwidth}{(a)}
          \fig{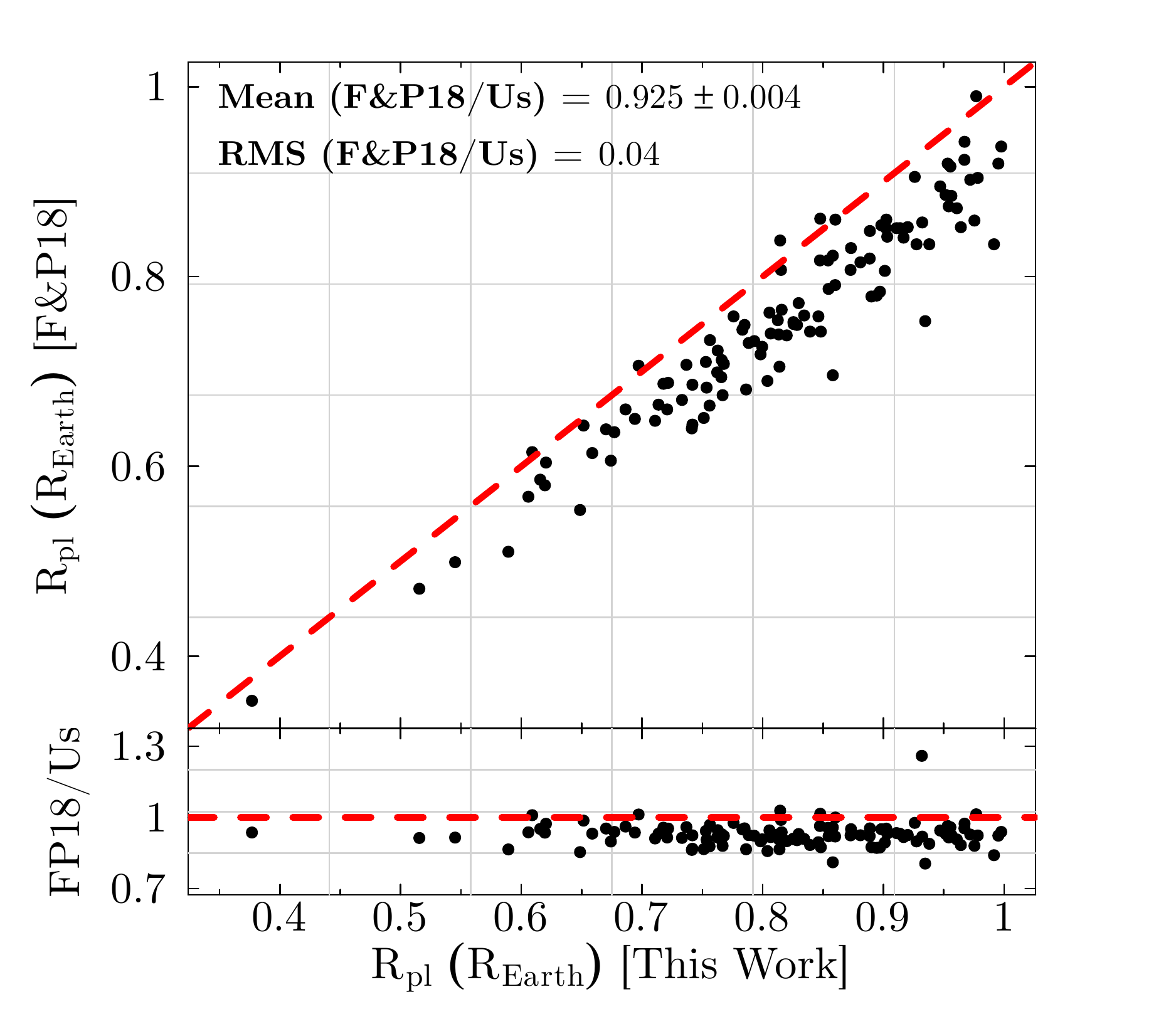}{0.52\textwidth}{(b)}
          }
\caption{Panel (a) presents the comparison of our planetary radii with those derived by \cite{fulton2018} for planets in common having planetary radii less than 200 R$_{\oplus}$ in  \cite{fulton2018}.
Panel (b) shows the same comparison shown in panel (a), but for planetary radii $\leq$ 1R$_{\oplus}$. The offset between the results is clear. The mean ratio between the radii in the studies and the corresponding RMS scatter are also shown.
The red dashed lines represent the equality line.}
\label{fig:pl}
\end{figure*}

Figure \ref{fig:pl} shows the comparison of the planetary radii. The results show good agreement but there are some outliers for which \cite{fulton2018} compute unrealistically large planetary radii when compared to ours (R$_{pl}$ $>$ 23R$_{\oplus}$ in \cite{fulton2018}). 
In addition, \cite{fulton2018} have a few planets with R$_{pl}$ ranging roughly between 1,000 R$_{\oplus}$ and 24,000 R$_{\oplus}$ (e.g., for K03891.01; R$_{pl}$ = 23,732 R$_{\oplus}$; planets with R$_{pl}>$200R$_{\oplus}$ are off scale in Figure \ref{fig:pl}a). 
It is worth noting that the errors associated with these unrealistic radii in \cite{fulton2018} are as large as the planetary radii themselves. 
We note that the discrepant planetary radii shown in the comparison of the results in Figure \ref{fig:pl} are not due to differences in the stellar radii, as the comparison between our respective stellar radii values agrees quite well.

If we remove the outliers from the comparison, or, remove from the \cite{fulton2018} sample those planets with R$_{pl}$ $>$ 23R$_{\oplus}$, we obtain 0.957 $\pm$ 0.007 for the mean ratio ``F$\&$P18/Us'' and a RMS of 3$\%$.
It is also found that the differences in planetary radii between this study and those from \cite{fulton2018} increases somewhat for the smallest planets (R$_{pl}<$1R$_{\oplus}$, with a mean planetary ratio of 0.925, where  Figure \ref{fig:pl}b) shows the comparison of our planetary radii $\leq$ 1R$_{\oplus}$ with those derived by \cite{fulton2018} for planets in common.


\begin{figure*}
\gridline{\fig{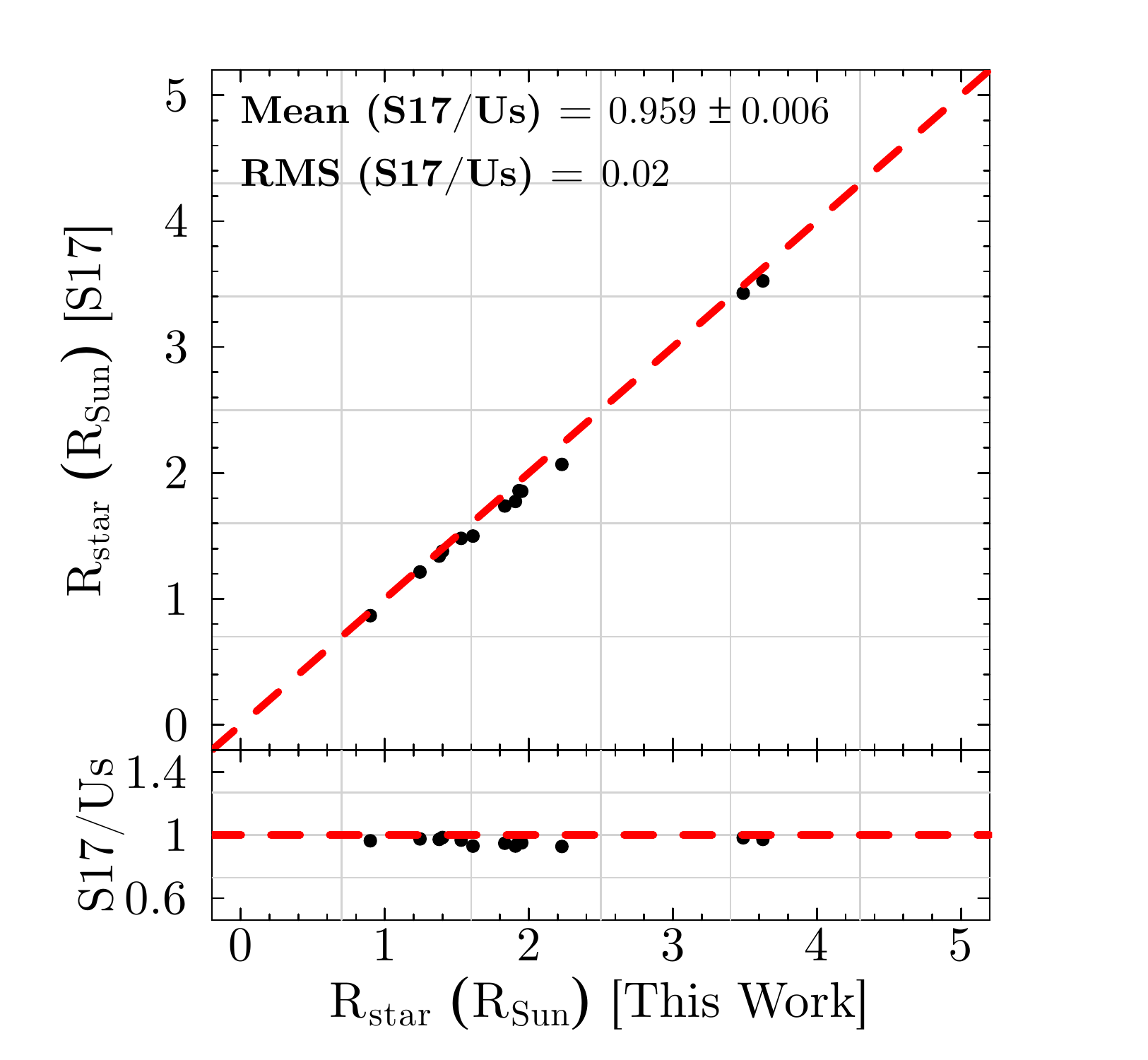}{0.36\textwidth}{(a)}
          \fig{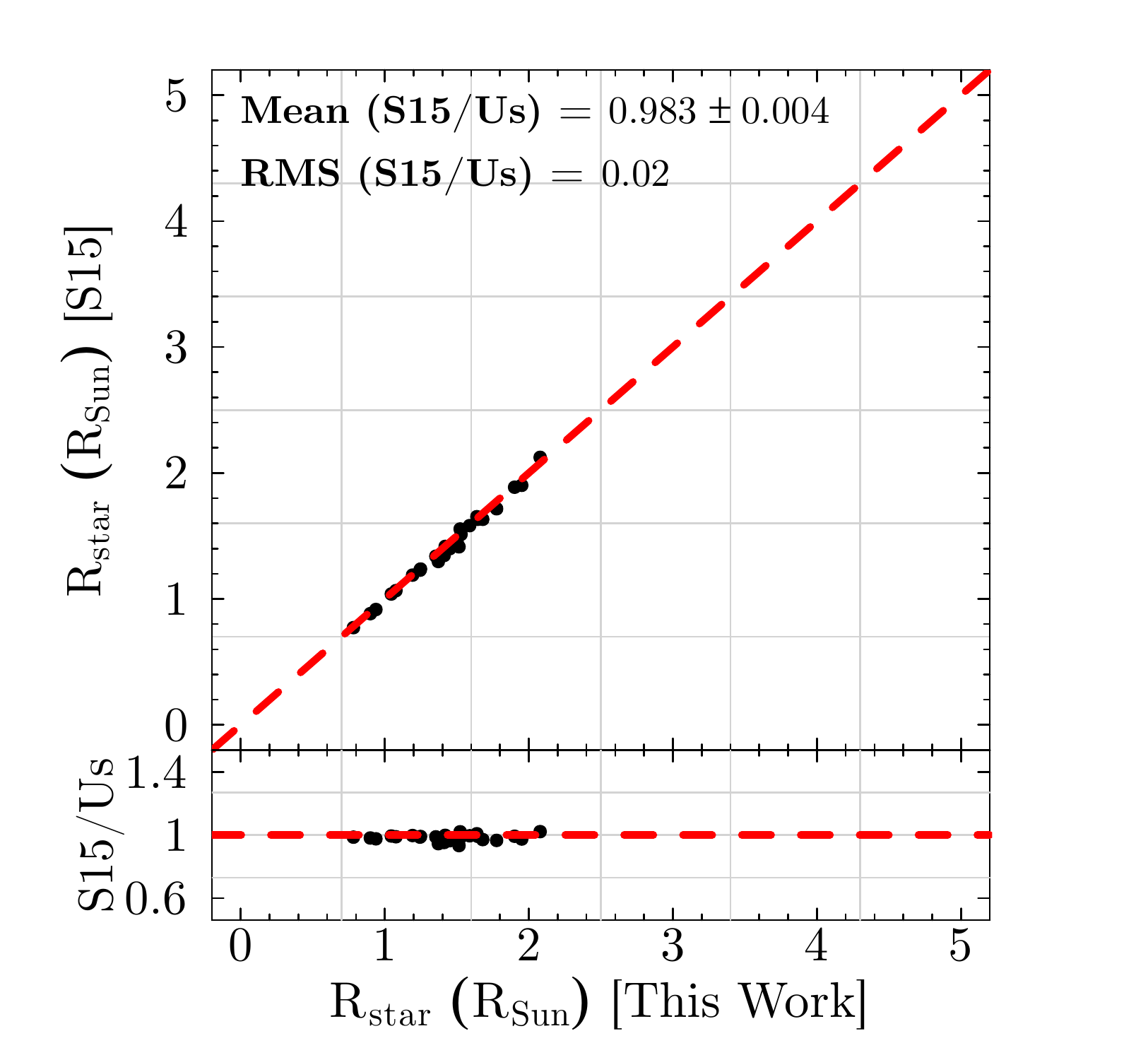}{0.36\textwidth}{(b)}
          \fig{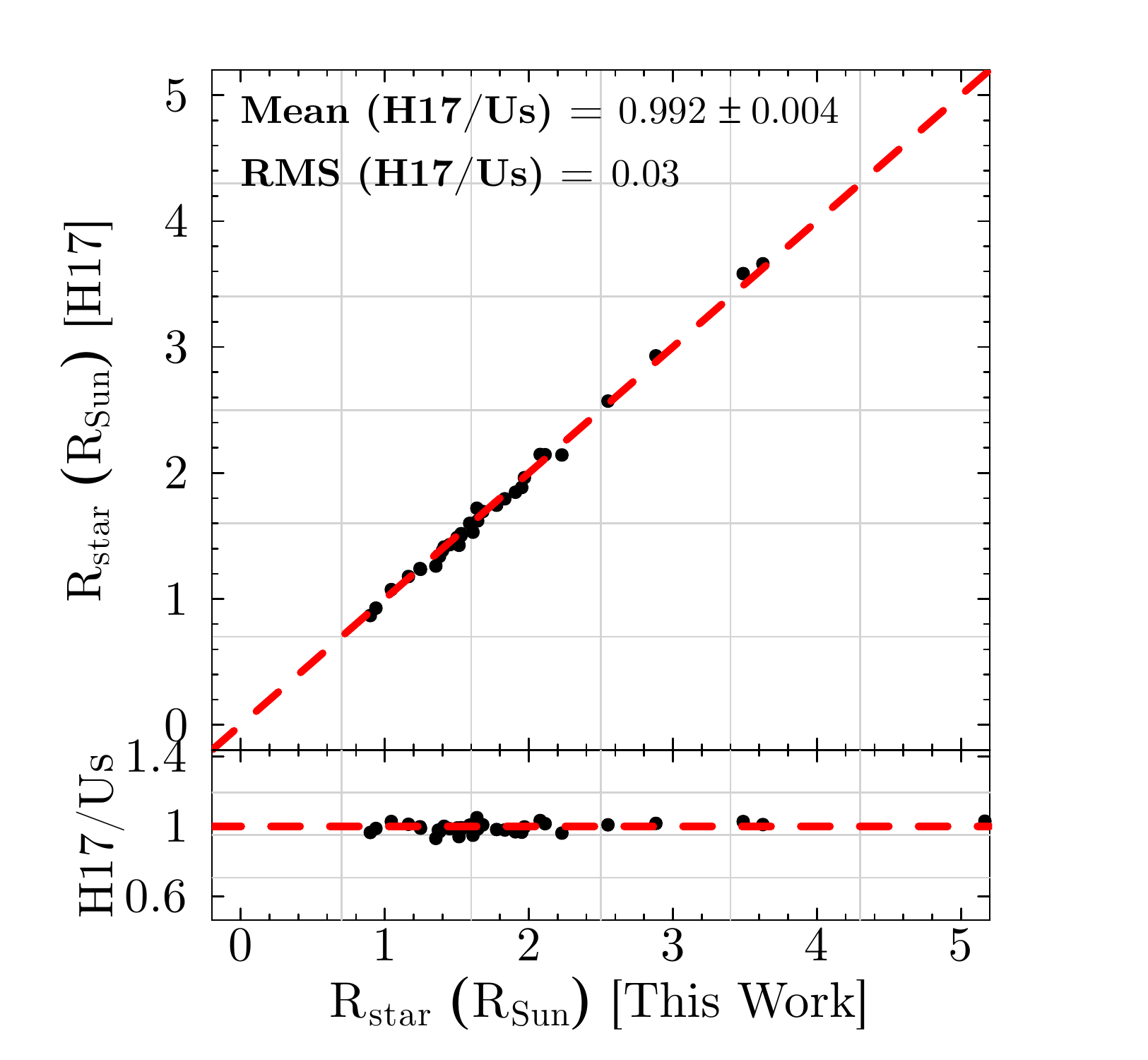}{0.36\textwidth}{(c)}
          }
\caption{A comparison of the derived stellar radii in this work against the precise stellar radii obtained from asteroseismology from (a) \citet{serenelli2017}, (b) \citet{silva-aguirre2015} and (c) \citet{huber2017}.}
\label{fig:sr-ast}
\end{figure*}


Figure \ref{fig:sr-ast} shows comparisons of the stellar radii derived in this study with results from asteroseismology. Such comparisons are particularly valuable because asteroseismology can deliver precise stellar radii, providing the basis for an assessment of both the accuracy and precision of our derived stellar radii. 
The three panels in Figure \ref{fig:sr-ast} show results for samples of stars in common with three asteroseismic studies (these same studies were used for the log $g$ comparisons in Figure \ref{fig:asteros}): \citet[panel a]{serenelli2017}, \citet[panel b]{silva-aguirre2015}, and \citet[panel c]{huber2017}. The mean values for the ratios ``Other studies/This study'', along with the corresponding RMS, are presented in each panel of Figure \ref{fig:sr-ast}; it is clear that our stellar radii compare very well with those from asteroseismology. 

It should be kept in mind, however, that the selected asteroseismic studies use di\-ffe\-rent methodologies (scaling relations versus modeling of individual frequencies), as well as different sets of stellar parameters (effective temperatures and metallicities), which are needed for their determinations of stellar radii. 
The internal precisions estimated for their derived radii are: 2.7$\%$ for \cite{serenelli2017}, 1.2$\%$ for \cite{silva-aguirre2015} and 2.7$\%$ for \cite{huber2017}. 
The internal precision inferred from our determinations of stellar radii (Table \ref{t:error}) is also at a similar level: $\sim$2.6$\%$. 
Concerning systematic offsets, we find that our derived radii are just slightly higher that the asteroseismic ones being on average 4.1$\%$ larger than those in \cite{serenelli2017}; 1.7$\%$  larger than those in \cite{silva-aguirre2015} and 0.8$\%$ larger than those in \cite{huber2017}; the RMS values obtained for the three studies are also quite small, indicating a small scatter of 0.02 -- 0.03. All in all, we can conclude that our derived stellar radii from precise parallaxes and precise spectroscopic determinations of the effective temperatures achieve a comparable precision against stellar radii derived from asteroseismology and do not contain systematic offsets that are much larger than typical offsets seen between the different asteroseismic studies themselves.

\section{Discussion}

\subsection{Exoplanetary Radii, Orbital Periods $\&$ Incident Fluxes}

\subsubsection{The Small Planet Radius Gap} \label{sec:gap}

Until recently, the detailed properties for the unprecedented variety of planets discovered by the $Kepler$ mission \citep{borucki2010,koch2010,borucki2016} have been ambiguous due in part to uncertainties in the planetary radii that stem from uncertainties in the stellar radii, which have been for the most part estimated from broad-band photometry.

Based on improved measurements for the stellar radii for the CKS sample by \cite{johnson2017} (using stellar and planetary parameters from \citealt{petigura2017} and \citealt{johnson2017}), \cite{fulton2017} found a bimodal distribution for the small planet radii having peaks at $\sim$1.3R$_{\oplus}$ and $\sim$2.4R$_{\oplus}$, with a gap between 1.5R$_{\oplus}$ - 2.0R$_{\oplus}$. 
\cite{vaneylen2018}, using asteroseismic radii for a small subsample of CKS targets with very precise radii, also detected a bimodal distribution with a clear gap around 2R$_{\oplus}$.
A similar gap in pla\-ne\-ta\-ry radius was confirmed by \cite{fulton2018} using $Gaia$ DR2 parallaxes. 
The dearth of planets at R$_{pl}$ $\sim$ 1.8R$_{\oplus}$ has been predicted by theoretical models and is interpreted as a transition radius separating planets with masses large enough to retain their gas envelope and those that have lost their atmospheres and consist of their remnant cores \citep{owen2013,lopez2014,chen2016,lopez2016,owen2017,ginzburg2018}. 
 
The distribution of planet radii derived in this study is shown in the different histograms in Figure \ref{fig:gap}. 
Although the methodology used in the determination of stellar parameters (needed to derive the stellar and planetary radii) for the CKS sample is completely independent of previous studies in the literature analyzing the same data set, the planetary radii distributions (presented in all panels of Figure \ref{fig:gap}) are found to be bimodal, showing the clear presence of a valley in the small-sized planet radius distribution. 
These independent results give support to the fact that the lack of planets around 2.0R$_{\oplus}$ is not an analysis artifact, but represents a real transition between rocky planets and those with extensive atmospheres \citep{weiss2014,rogers2015}.

The full sample analyzed in this study (composed of 1633 planets), without any cuts, is shown in Panel (a) of Figure \ref{fig:gap}. 
It should be kept in mind, however, that the full sample does not include KOI's with planets deemed as ``false positives'' (see Section \ref{sec:obser}).
The uncorrected-completeness radius distribution shows two clear peaks: one at $\sim$1.6R$_{\oplus}$ and another at $\sim$2.8R$_{\oplus}$; the radius gap is seen roughly between 1.8R$_{\oplus}$ and 2.2R$_{\oplus}$.

It is of interest to investigate the position of the peaks and the planetary radius gap using only precise planetary radii, as it is clear that the derived radii have different levels of precision depending, for example, on the errors in the $Gaia$ DR2 parallaxes, which are folded into the distance error estimates by \cite{bailer2018}. 
In panel (b) we show a similar histogram as in panel (a), but in this case we restricted our sample to consider only those planets with uncertainties in the derived radii of less than 8$\%$ (corresponding to $\sim$2$\times$ the median uncertainty; Section \ref{sec:errors}).
In panel (c) we applied the same selection as in panel (b), but similarly to \cite{fulton2017}, we removed from the sample any planets with $P$ $>$ 100 days and in a transit configuration corresponding to the impact parameter, $b$, being larger than 0.7; this is our ``clean'' sample, which is composed of 965 planets.

The radii histograms shown in Figure \ref{fig:gap} panels (a) (for the entire sample), (b) (for planets with precise radii) and (c) (for the ``clean'' sample) are in general quite similar: in all distributions there is a dearth of planets at $\sim$2.0R$_{\oplus}$ and the location of the two peaks in the distributions is similar as well. 
The full sample has a small excess of planets at R$_{pl}$ $\sim$ 10R$_{\oplus}$ and this is mostly due to the presence of planets with $P$ $>$ 100 days that were removed in the ``clean'' sample. 
There is also a large population of small planets with R$_{pl}$ $<$ 1.0R$_{\oplus}$, but as pointed out by \cite{berger2018}, it 
is expected that some of these small planets will be classified as ``false positives'' in the future. 

\subsubsection{Completeness Corrections} 

To assure that the trend described in the previous Section is not an artifact of completeness and  affected by the lack of detectability of planets with small radii and/or long orbital periods by $Kepler$, we reconstruct the planet occurrence rate of the $Kepler$ sample after applying completeness corrections \citet{fulton2018,fulton2017, mulders2018, mulders2016, christiansen2016, christiansen2015}.

We used the injection-recovery experiments described in \cite{christiansen2015,christiansen2016}. They measured the $Kepler$ pipeline detection efficiency by injecting simulated transiting planets into the raw pixel data and analyzed the recovery rate after processing them with the $Kepler$ pipeline, to reconstruct the planet occurrence rate of the $Kepler$ sample. 

Considering injections in the transit signals for the stars included in our sample, we used our derived stellar and planetary radii to calculate the reliability of the transit measure ($m$) following the procedure described in \cite{fulton2017}:

\begin{displaymath}
     m = (\frac{R_{pl}}{R_{\star}})^2 \frac{1}{CDPP_{dur}} \sqrt{\frac{t_{obs}}{P}}
\end{displaymath}

\noindent where t$_{obs}$ is the time that a star of radius $R_{\star}$, harboring a planet of radius $R_{pl}$ and orbital period $P$, was observed; while the Combined Differential Photometric Precision \cite[$CDPP_{dur}$,][]{koch2010} is the noise of a transit signal interpolated to the transit duration time. The transit  parameters necessary to do the calculations were taken from the $Kepler$ database \cite[DR25,][]{thompson2018}.

To account for the pipeline efficiency, \cite{fulton2017} fit a $\Gamma$ cumulative distribution function of the form

\begin{displaymath}
 C(m;k,\theta,l) = \int_{0}^{\frac{m-l}{\theta}} x^{k-1} e^{-x}\cdot dx
\end{displaymath}

\noindent to model the distribution of values from the injection-recovery transit signals ($m$); we use the $k$, $l$ and $\theta$ values as determined by \cite{fulton2017}.

We determined the detection probability, $p_{det}$, obtained by using $C(m)$ values, and the geometric transit probability, $p_{tr}$, to check for the survey sensitivity.

The transit probability, $p_{tr}$, defined as the geometric probability of a planet with radius R$_{pl}$ transiting a host star of radius R$_{\star}$ at a distance of $a$ could be detected, is $p_{tr}$ = 0.7R$_{\star}$/$a$; where the factor 0.7 corresponds to the imposed limit in the impact parameter of the planet candidates, according to the work of \citeauthor{fulton2017}.

To compensate for incompleteness due to a lack of detection efficiency, $p_{det}$, or a low probability of transit detection, $p_{tr}$, we weighted each planet by the inverse of these probabilities:

\begin{displaymath}
     w_{i} = \frac{1}{p_{det}\times p_{tr}}  
\end{displaymath}

The true measure of the occurrence rate, $f_{bin}$, is the number of planets per star in any orbital period or radius bin, is then given by:

\begin{displaymath}
     f_{bin} = \frac{1}{N_{\star}} \sum_{i=1}^{n_{pl,bin}}w_{i} 
\end{displaymath}

\begin{figure}
\epsscale{0.85}
\plotone{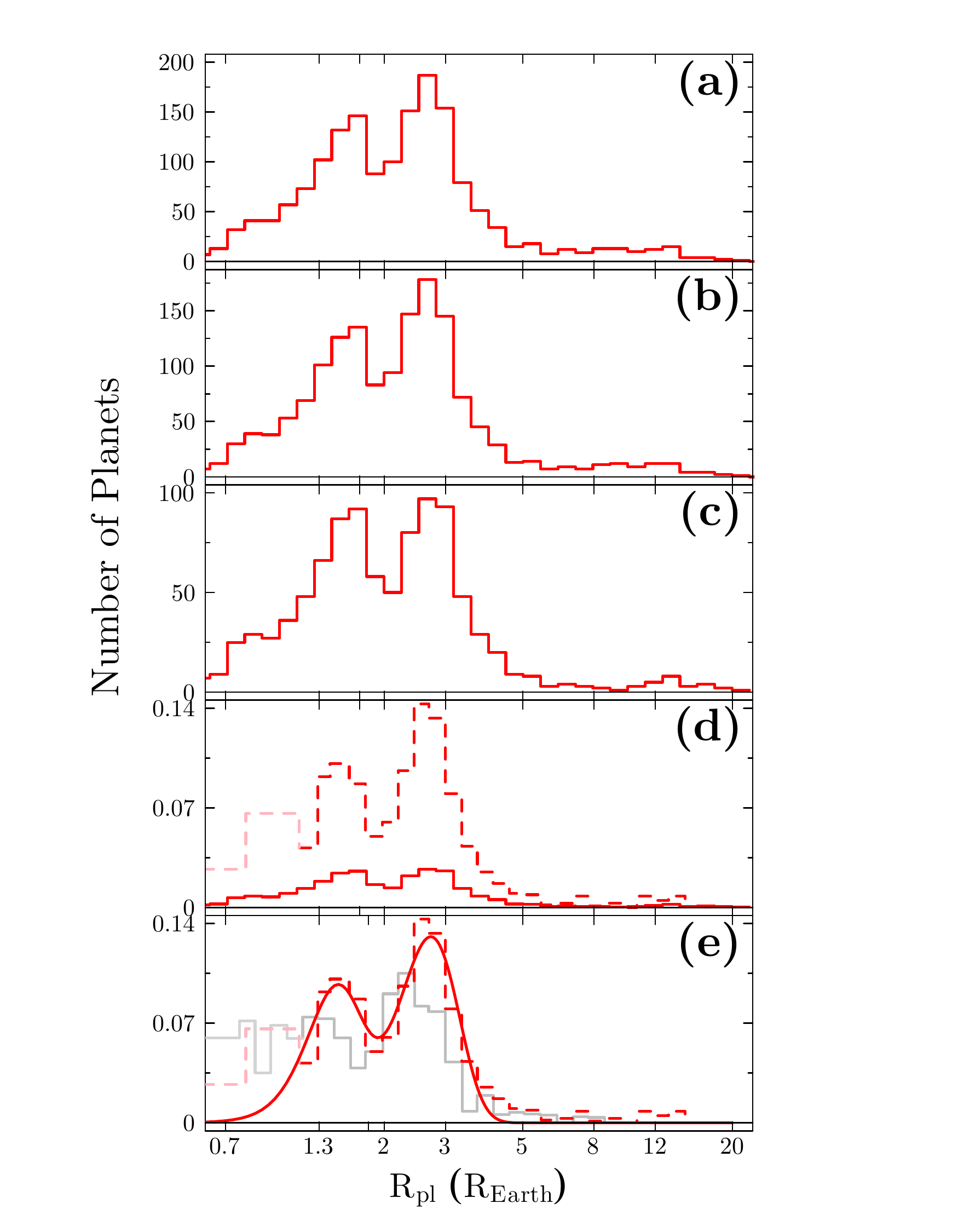}
\caption{Radius distribution (a) considering the entire planet sample (1633 planets); (b) taking into account only those planets (1526 planets) having radii with errors less than 8$\%$; (c) the same as panel (b) but also discarding planets with $P$ $>$ 100 days and planetary systems with $b$ $>$ 0.7 (this defines the ``clean'' sample having 965 planets); (d) completeness-corrected histogram of planet radii for our ``clean'' sample (dashed line) in comparison with the distribution of planet radii uncorrected for completeness (solid line); (e) completeness-corrected planet radii for our ``clean'' sample (red line) but compared with the one from \citet{fulton2018} (grey line). We confirm the presence of a gap in the occurrence distribution of close-in planets with orbital periods less than 100 days at $\sim$1.9R$_{\oplus}$.}
\label{fig:gap}
\end{figure}

Figure \ref{fig:gap}(d) presents the same distribution of planetary radii for our ``clean sample'' (shown in panel (c)) in comparison with the completeness-corrected distribution, shown as the dashed line histogram. It is noticeable that the location of the gap and the peaks in the completeness-corrected distribution are shifted to slightly smaller radii when compared with the uncorrected distribution, similarly to what has been found by \cite{fulton2018}.

The location of the peaks of the completeness-corrected planetary radius distribution can be estimated from the kernel density estimate for a Gaussian distribution (shown in Figure \ref{fig:gap}(e)); the peak positions of the distribution are found at 1.47 $\pm$ 0.05 R$_{\oplus}$ and 2.72 $\pm$ 0.10 R$_{\oplus}$, and a radius gap at 1.89 $\pm$ 0.07 R$_{\oplus}$ (Figure \ref{fig:gap}(e)). The completeness-corrected distribution from \cite{fulton2018} is also shown for comparison as the grey line histogram in Figure \ref{fig:gap}(e). There is a marginal shift in our planetary distribution relative to \cite{fulton2018}.
\cite{vaneylen2018} also investigated the planetary radii distributions in their, albeit smaller sample, but with precise radii from asteroseismic parameters; they find the peaks to be at 1.5R$_{\oplus}$ and 2.5R$_{\oplus}$, respectively, with the radius gap minimum falling at 2.0R$_{\oplus}$.


\subsubsection{A Slope in the Planetary Radius Gap} \label{sec:slope}

As discussed in Section \ref{sec:gap}, one mechanism suggested to explain the bimodal small planet radius distribution is photo-evaporation: X-ray and UV fluxes from the young planet-hosting star evaporate the envelopes of the H-He rich sub-Neptunes, exposing their stripped rocky cores. In addition to photo-evaporation, \cite{ginzburg2018} model the radius valley as being caused by the energy from young, hot planetary cores driving planetary-mass dependent atmospheric mass loss (or ``core-powered mass-loss''). Several theoretical models also predict the shape and the slope of the evaporation valley \citep[e.g.,][]{owen2013,owen2017,jin2018}; the planetary radii for the CKS derived in this study are precise enough to investigate the presence of such signatures in the planetary radius as a function of orbital period or incident flux.

Figures \ref{fig:size} and \ref{fig:flux} present the derived exoplanetary radii as a function of their orbital periods (with $Kepler$ DR25 periods taken from \citealt{thompson2018}) and insolation fluxes, respectively. 
The stellar fluxes were calculated from the derived T$_{\rm eff}$, and R$_{\star}$, while the semi-major axis of the planetary orbits, assuming an orbital eccentricity equal to zero, were used to compute the incident flux at the planet. 

In both figures, panel (a) shows the full sample with 1633 planets, without any cuts, and panel (b) shows a subset of the ``clean'' sample for the region containing the smaller exoplanets (radii less than $\sim$ 4R$_{\oplus}$). 
Visual inspection of these figures indicates clearly the presence of two clouds representing the population density of the distributions of super-Earths and sub-Neptunes in the log Radius - log Period/Incident Flux planes, with a low density of exoplanets at radii in-between the two clouds, with R$_{pl}$ $\sim$ 2R$_{\oplus}$. 
This ``evaporation valley'' corresponds to the separation between the super-Earth and sub-Neptune regimes. 

The exoplanet population plotted in Figures \ref{fig:size} and \ref{fig:flux} also shows an overall lack of sub-Neptunes with short orbital periods (P $<$ $\sim$ 3 days) and incident fluxes (relative to Earth) $>$ $\sim$ 700, which is likely related to the photo-evaporation of the atmospheres of sub-Neptune-size exoplanets that are very close to their parent stars and suffer high stellar incident flux levels \citep{ikoma2012,lopez2013,ciardi2013,owen2013,wu2013}. 
It should be pointed out, however, that some exoplanets have been detected in the sub-Neptune ``desert'' region \citep[see][]{west2018}.

Close inspection of Figure \ref{fig:size} gives some hints that the value of R$_{pl}$ in the evaporation valley minimum overall decreases with increasing orbital period.  
The change in the radius gap minimum as a function of orbital period can be quantified in a simple way by dividing our ``clean'' sample of planets (with radii errors $\leq$ 8$\%$, $P$ $\leq$ 100 days and $b$ $\leq$ 0.7) into ten orbital period bins containing equal number of planets. 
Within each orbital period bin, the minimum value in the radius gap was measured resulting in a linear relation defined by log(R$_{pl}$) = (-0.11 $\pm$ 0.02)*log($P$) + (0.39 $\pm$ 0.01), with R$_{pl}$ in units of Earth radii and P in days, or R$_{pl}$ scaling as $P^{-0.11}$; the slope along with the corresponding prediction interval is shown in Figure \ref{fig:size}(b)).
A similar fit was done for the valley observed between the planetary radii and the incident fluxes (Figure \ref{fig:flux}(b), finding a linear relation defined as: 
log(R$_{pl}$) = (+0.12 $\pm$ 0.02)*log($F$) + (0.04 $\pm$ 0.03), with F in units of incident flux at the Earth, or R$_{pl}$ scaling as \textbf{$F^{+0.12}$}. 
 
\cite{vaneylen2018} also detected a slope in the planetary radius valley as a function of the planetary orbital period from a small sample of 75 stars, well characterized by asteroseismology, and having precise radii for their 117 associated planets. 
Their derived slope of $P^{-0.09^{+0.02}_{-0.04}}$ for the small planet radius gap, although from a much smaller sample, is in good agreement with ours, within the uncertainties. 
They also model the valley slope after restricting their sample to periods $<$ 25 days and find $P^{-0.10}$; if we follow the same cut for the much larger CKS sample we obtain a similar slope of $P^{-0.11 \pm 0.03}$.

The shape and, in particular, the value of the slope in the evaporation valley can constrain planetary formation models. \cite{lopez2016} used different models to point out that the transition radius between rocky super-Earths and sub-Neptunes with volatile envelopes scales differently with the orbital period depending on the planet formation scenario; the transition radius should decrease for longer orbital periods in the case of a photo-evaporation scenario (R$_{pl}$ scaling as $P^{-0.15}$). 
Contrarily, it should increase if the primordial rocky planets formed after the proto-planetary disks dissipated in a gas-poor formation model with a positive slope for the radius gap (R$_{pl}$ scaling as $P^{+0.07 \pm 0.10}$). 

According to models by \cite{owen2017}, the slope of the transition radius (or the upper envelope of the super-Earth radius) with period derived through evaporation models can change when considering different evaporation efficiencies in these models, ranging from $P^{-0.25}$, when a constant evaporation efficiency is considered for all the planets, to $P^{-0.16}$, when evaporation efficiency depends on the planet density.  
\cite{owen2017}, as well as \cite{jin2018}, also investigate how the bulk composition affects the planetary radii of super-Earths and sub-Neptunes as a function of period (or orbital semi-major axis).  
Depending on the mixture of iron, silicates, or ices, for example,
the maximum radius of super-Earths changes as a function of orbital period.
Using Figure 10 from \cite{owen2017} as an example, their model which
mimics the composition of the Earth (with $\sim$1/3 iron) and considers variable efficiency for the evaporation, provides the closest match to the upper envelope de\-fi\-ning super-Earth radii versus period, for periods less than $\sim$8 days.  
The model then declines more steeply than the observed distribution towards increasing periods, with the result that the observed maximum radius for super-Earths falls well above the \cite{owen2017} model core with 1/3 iron.  
Since their ``icy'' models with 1/3 ice and 2/3 silicates have larger radii than the iron cores, the observed super-Earths with larger radii at longer periods may signal a shift in the overall compositions of super-Earths with short periods compared to those with long periods.

The location of the radius gap in Figure \ref{fig:size} and its negative slope agree with what is expected from photo-evaporation models.  
The change in the maximum radius versus orbital period observed for super-Earths in Figure \ref{fig:size} agrees qualitatively with model cores from \cite{owen2017}; a quantitative comparison with model predictions can constrain the core compositions and may even be able to map compositions at different orbital periods around different types of host stars.

The shape and location of the planet-radius versus orbital period valley is also in qualitative agreement with the core-powered mass-loss mechanism from \cite{ginzburg2018}. In this model, atmospheric mass loss is more effective if the equilibrium temperature of the planet, T$_{\rm eq}$, is high, which is, in general the case for more closely orbiting planets. \cite{ginzburg2018} evolve a model planetary distribution and find that the position of the valley minimum shifts from $\sim$2R$_{\oplus}$ for P$<$10 days, to $\sim$1.5R$_{\oplus}$ for the longer orbital periods. Recently, \cite{fulton2018} have es\-ta\-blished a relation between the cumulative distribution of planets versus the incident flux as a function of stellar mass, with the distribution shifted to higher incident fluxes for larger stellar masses. The core-powered mass-loss mechanism is dependent on the properties of the planet itself, with no expected dependence on the stellar mass; the host-star mass correlation with the planetary incident flux distribution may favor the photo-evaporation model.  

Figure \ref{fig:flux}, where planetary radii are plotted versus incident flux, tells effectively the same story as Figure \ref{fig:size}. The incident flux shown is the current flux, although within the photo-evaporation model, \cite{owen2013,owen2017} point out that it is the
flux from the young, presumably active host star that is most important
in sculpting the distributions of planetary radii in Figures \ref{fig:size} and \ref{fig:flux}. Photo-evaporation is most effectively driven by the stellar integrated X-ray and extreme UV (EUV) fluxes during the first 100 Myr of the life of the star \citep{owen2013}.  
The planetary composition also plays a role, as H and He are most affected by the EUV flux, while the metals are more easily evaporated by the X-ray flux \citep{owen2017}.  
The planetary radii distributions as functions of orbital period and incident flux shown in Figures \ref{fig:size} and \ref{fig:flux} are thus the result of their early X-ray and EUV radiation environments and distance from their young host star.

\subsubsection{A Possible Correlation between Planetary Radii and Orbital Periods?} \label{sec:flux}

The CKS planetary sample studied here can be divided into four exoplanet size regimes:

\begin{itemize}
\item Jupiters with 8R$_{\oplus}$ $<$ R$_{pl}$ $\leq$ 20R$_{\oplus}$. 
\item Sub-Saturns with 4R$_{\oplus}$ $<$ R$_{pl}$ $\leq$ 8R$_{\oplus}$
\item Sub-Neptunes with 2R$_{\oplus}$ $<$ R$_{pl}$ $\leq$ 4R$_{\oplus}$
\item Super-Earths with R$_{pl}$ $\leq$ 2R$_{\oplus}$
\end{itemize}

The sample (in this case we are considering the full sample of planets without any cuts) has a majority of exoplanets with small sizes, split roughly in equal numbers between the classes of super-Earths (736 exoplanets) and sub-Neptunes (706 exoplanets). 
It has a much smaller number of sub-Saturns (96 exoplanets) and 93 exoplanetary systems containing at least one Jupiter-like planet; about 20$\%$ of these systems have a single Hot Jupiter without any detectable inner or outer companions, while almost half of the Warm Jupiter sample ($\sim$30 out of $\sim$70) is found around a variety of exoplanetary system architectures. We define Hot Jupiters as those with periods less than 10 days, while Warm Jupiters have periods larger than 10 days.  

\begin{figure*}
\gridline{\fig{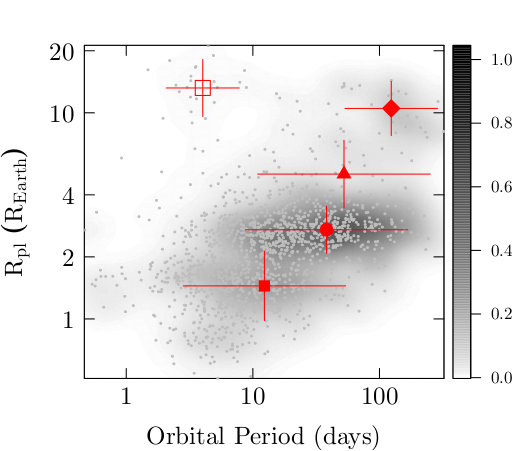}{0.55\textwidth}{(a)}
          \fig{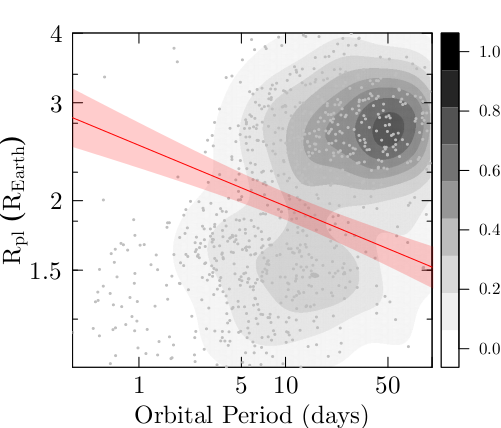}{0.55\textwidth}{(b)}
          }
\caption{(a) Planetary radius as a function of planetary period for the entire sample. Red symbols and bars indicate the weighted median values and their weighted uncertainties, respectively, for planetary radii and orbital periods for each planet size domain: super-Earths (filled square), sub-Neptunes (circle), sub-Saturns (triangle), Hot Jupiters (open square) and Warm Jupiters (diamond).
(b) Same as (a) but only considering the small-sized planet regime (radii less than $\sim$ 4R$_{\oplus}$) of our ``clean'' planet sample. The best fit slope (red line) and prediction interval (shaded region) to the radius gap is shown.}
\label{fig:size}
\end{figure*}

\begin{figure*}
\gridline{\fig{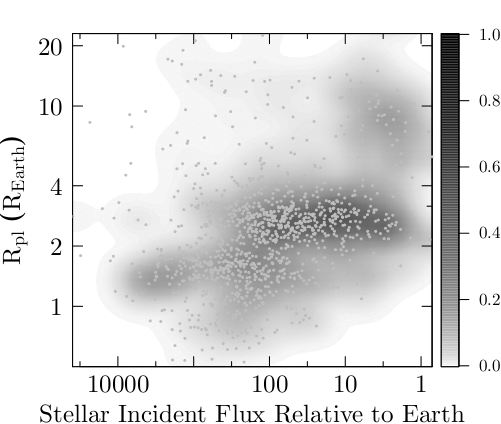}{0.55\textwidth}{(a)}
          \fig{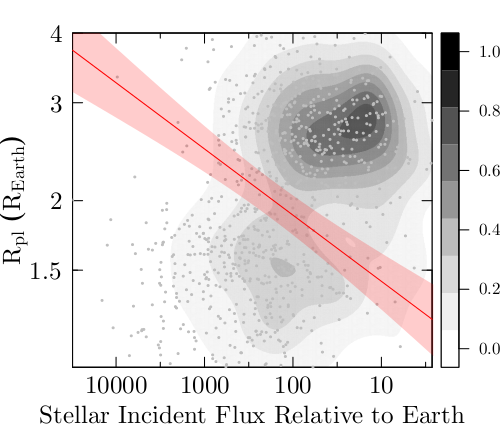}{0.55\textwidth}{(b)}
          }
\caption{Panel (a): Planetary radius as a function of stellar incident flux (relative to Earth) for the entire planet sample. Panel (b): Same as (a) but only considering the small-sized planet regime (radii less than $\sim$ 4R$_{\oplus}$) of our ``clean'' planet sample. The best fit slope (red line) and prediction interval (shaded region), to the photo-evaporation valley is also shown.}
\label{fig:flux}
\end{figure*}

Figure \ref{fig:size}(a) also shows, as red symbols, the weighted median values and weighted RMS scatter of the planetary radii and orbital periods distributions for each planet size domain: super-Earths (filled square), sub-Neptunes (circle), sub-Saturns (triangle), Hot Jupiters (open square) and Warm Jupiters (diamond).
It is clear that the super-Earths have a median value for planetary radii and orbital periods that is lower than the sub-Neptunes, suggesting a possible correlation in the sense that the sizes of the exoplanets generally increase with orbital periods; some of this correlation is due to incompleteness in transit measurements given, for example, the difficulty in detecting small planets at larger distances from the parent star (with larger orbital periods). 
The correlation between planetary radii and orbital periods extends, however, towards the larger exoplanet groups containing the sub-Saturns and Warm Jupiters, for these larger planets the observational biases should not be significant, in particular for systems with orbital periods less than $\sim$ 500 days (taking completeness values from \citealt{silburt2015}). 

The Hot Jupiters do not fit into the trend of increasing orbital period with increasing exoplanet size. 
The median value of R$_{pl}$ - $P$ for the Warm Jupiters (diamond) generally follows the trend delineated by the smaller planets, while the Hot Jupiters occupy a different locus in the R$_{pl}$ - orbital period plane (median represented by the open square), having much shorter orbital periods on average and being much closer to their parent stars.
Hot Jupiters are believed to have formed several AUs away from their parent stars and undergone extreme migration, into the exoplanetary system's inner region, destabilizing any small exoplanets, scattering them out of the system, or destroying them \citep{latham2011,morbidelli2014,mustill2015}.  
We also note that \cite{huang2016} analyzed a sample of 45 Hot Jupiters and 27 Warm Jupiters from the $Kepler$ catalog and proposed that Warm Jupiters with low-mass companions would be formed \textit{in-situ}, not affecting their small neighbors, and that Warm Jupiters with no detectable companions may be a distinct population.


With the position of the Hot Jupiters in the R$_{pl}$ - $P$ plane dominated by migration, the other size groups (super-Earths, sub-Neptunes, sub-Saturns, and Warm Jupiters) may define a seemingly tight correlation of median radius with median orbital period. This correlation is influenced by the dearth of sub-Neptunes with short orbital periods. 

To evaluate how much the correlation depends on the limits adopted for cutting the planet sample in terms of radii and orbital periods, we analyze three different samples that will correspond to different levels of completeness.
1) We consider the entire sample of Warm Jupiters, Sub-Saturns, Sub-Neptunes and Super-Earths, at any orbital period, without any considerations about observational biases. A fit to the trend in log-log space results in a well-defined power-law relation, such that R$_{pl}$ $\sim$ $P^{0.84 \pm 0.11}$, with R$_{pl}$ in Earth-radii and orbital period in days. 
2) If we limit the sample to include only those planets with R$_{pl}$ $\geq$ 2R$_{\oplus}$ (equivalent to assuming that roughly all Sub-Neptunes and larger planets can be detected at any orbital period) we obtain a steeper power-law but with smaller errors in the fit: R$_{pl}$ $\sim$ $P^{1.1 \pm 0.3}$. 3) If we add now a cut in orbital period, considering only those planets with R$_{pl}$ $\geq$ 2R$_{\oplus}$ and with $P$ $<$ 500 days, we obtain he same power-law as before but with an increase in the uncertainties in the fit (R$_{pl}$ $\sim$ $P^{1.09 \pm 0.13}$). 
We expect that the $Kepler$ completion levels would be significant for this regime and verify from these tests that the correlation of planetary radii with orbital periods does not change significantly in the 3 samples analyzed.

\cite{helled2016} also suggested the existence of a correlation between planetary radius and orbital periods for exoplanets with radii smaller than 4R$_{\oplus}$ using $Kepler$ data. 
They did a statistical analysis that took into consideration completeness values for the detection of planets by $Kepler$ \citep{silburt2015} to conclude that this correlation was not the result of a selection bias. \cite{helled2016} obtained a power law relation between planet radius and orbital period of: R$_{pl}$ $\sim$ $P^{0.5 - 0.6}$, which is similar, within the uncertainties, to the value of $\sim$0.8 obtained here. 
If true, the correlation between radii and orbital periods found for the smaller planets in the CKS sample may imply that larger planets would also more likely form at larger distances from the host star.

\section{Conclusions}

$\bullet$ We have conducted a homogeneous, quantitative spectroscopic analysis of 1232 exoplanet host stars using the high-resolution Keck/HIRES spectra made publicly available by the California-$Kepler$ Survey (CKS) team  \citep{petigura2017,johnson2017}. 
Stellar parameters (T$_{\rm eff}$ and log $g$) were derived from an equivalent width measurement analysis of a sample of 158 Fe I and 18 Fe II lines and using the automated pipeline described in \cite{ghezzi2010b,ghezzi2018}.  
\smallskip

$\bullet$ $Gaia$ DR2 parallaxes \citep{gaia2018} for the 
CKS stars were used to determine precise distances in \cite{bailer2018}, which were then used as the foundation for determining stellar luminosities in this study. With tightly constrained effective temperatures and luminosities, stellar radii were computed for the entire sample, with a median internal uncertainty of 2.8$\%$. 
\smallskip

$\bullet$ Our derived stellar radii from precise parallaxes and precise spectroscopic determinations of the effective temperatures achieve a comparable precision against stellar radii obtained from asteroseismology, with no significant systematic offsets. 
Precise stellar radii are important to constrain the planetary radii - a crucial parameter ne\-ce\-ssa\-ry to unveil planetary composition.
\smallskip

$\bullet$ Considering the sample of those stars with R$_{\star}$ uncertainties less than 10$\%$ and planetary transit depths from \cite{thompson2018}, we derive planetary radii for 1633 planets, 
with a median uncertainty of 3.7$\%$.
Comparisons of our derived planetary radii with those from \cite{fulton2018} indicate that our planetary radii are systematically larger than theirs by $\sim$ 4.3$\%$.
\smallskip

$\bullet$ The derived planetary radii clearly exhibit two peaks in the completeness-corrected planetary radii distributions. In particular, for our ``clean'' planet sample (planets with radii errors $<$ 8$\%$, $P$ $<$ 100 days and $b$ $<$ 0.7), we obtain peaks in the radius distributions corresponding to  1.47R$_{\oplus}$ (super-Earths) and 2.72R$_{\oplus}$ (sub-Neptunes), with a clean minimum (the ``gap'') at 1.9R$_{\oplus}$. 
\smallskip

$\bullet$ Given the good internal precision in the derived radii, it was possible to evaluate not only the location but the shape of the radius gap. Our results indicate that the radius gap for the CKS sample does not fall at a constant value of radius, but changes as functions of both planetary orbital period and incident stellar flux. The position of the radius gap decreases with orbital period and this decrease can be fit by a power law of the form R$_{pl}$ $\propto$ P$^{-0.11}$; this agrees well with the recent value of -0.09 from \cite{vaneylen2018} for a much reduced sample of 117 planets having precise radii from asteroseismology.  
\smallskip

$\bullet$ According to \cite{owen2017}, the value of the slope in the evaporation valley can constrain the planet core compositions at different orbital periods around different types of host stars. In our case, R$_{pl}$ $\propto$ P$^{-0.11}$ matches a terrestrial-like composition model for the planets in the transition radius.
\smallskip

$\bullet$ The value of the radius gap increases with increasing incident stellar flux, such that R$_{pl}$ $\propto$ F$^{+0.12}$ provides an excellent fit to our results. The trend of radius gap position with incident flux (and orbital period) agrees with models of photo-evaporation \citep[e.g.,][]{owen2013,owen2017,jin2018}.
\smallskip

$\bullet$ If we divide our planetary sample into Warm Jupiters (Jupiters with orbital pe\-riods larger than 10 days), Sub-Saturns, Sub-Neptunes and Super-Earths, we find that the mean values for planetary radii and orbital periods seem to suggest a po\-ssi\-ble correlation: larger planets seem to form more distant from their parent stars. 
Considering only those planets with radii $\geq$ 2R$_{\oplus}$ and with orbital periods $<$ 500 days (for which $Kepler$ completeness levels should be high), we obtain R$_{pl}$ $\propto$ $P^{0.8}$; this slope is similar to what has been found previously for small planets by \cite{helled2016}.

\clearpage

\acknowledgments

We thank the referee for giving suggestions that improved the paper. We warmly thank the California-$Kepler$ Survey team for making their reduced data publicly available. We thank the European Space Agency (ESA) mission $Gaia$ and NASA Exoplanet Archive for the data used in this work. CM acknowledges the financial support from Conselho Nacional de Desenvolvimento Cient\'ifico e Tecnol\'ogico (CNPq). CM also thanks D.F. Morell for his kindly assistance in R language. KC and VS acknowledge that their work is supported, in part, by the National Aeronautics and Space Administration under Grant 16-XRP16 2-0004, issued through the Astrophysics Division
of the Science Mission Directorate.
LG would like to thank the financial support from Coordena\c{c}\~{a}o de Aperfei\c{c}oamento de Pessoal de Nível Superior (CAPES).
\\


\software{Python packages: astropy \citep{astropy}, 
         $iso\-cla\-ssi\-fy$ \citep{huber2017} and R packages: astro \citep{astro-R}, mclust \citep{mclust-R}, MASS \citep{mass-R}.}


\end{document}